\documentclass[12pt,preprint2]{emulateapj}
\usepackage{pslatex}
\usepackage[T1]{fontenc}
\usepackage[latin1]{inputenc}
\usepackage{subfigure}
\usepackage{amsmath}
\usepackage{graphicx,graphics}
\usepackage{amssymb}
\usepackage[english]{babel}

\newcommand{\gap}{\;\rlap{\lower 2.5pt \hbox{$\sim$}}\raise 1.5pt\hbox{$>$}\;}
\newcommand{\lap}{\;\rlap{\lower 2.5pt \hbox{$\sim$}}\raise 1.5pt\hbox{$<$}\;}
\newcommand{\beq}{\begin{equation}}
\newcommand{\eeq}{\end{equation}}
\newcommand{\msun}{M_\odot}
\newcommand{\mh}{M_\bullet}

\newcommand{\tr}{T_{\rm r}}
\newcommand{\tgr}{T_{\rm gr}}
\newcommand{\tcoal}{t_{\rm coal}}
\newcommand{\thard}{T_{\rm hard}}

\newcommand{\mtwo}{M_{12}}
\newcommand{\aeq}{a_{\rm eq}}

\newcommand{\q}{\alpha}

\shorttitle{Binary Black Holes}
\shortauthors{Merritt, Mikkola \& Szell}

\begin{document}

\title{Long-Term Evolution of Massive Black Hole Binaries. 
III. \\
Binary Evolution in Collisional Nuclei}

\author{David Merritt}
\affil{Department of Physics, 85 Lomb Memorial Drive, Rochester Institute of Technology, Rochester, NY 14623\\ {\it and} \\
Center for Computational Relativity and Gravitation, School of Mathematical Sciences, 78 Lomb Memorial Drive, Rochester Institute of Technology, Rochester, NY 14623}


\author{Seppo Mikkola}
\affil{Turku University Observatory, Tuorla, 21500 Piikki\"o, Finland}


\author{Andras Szell}
\affil{Department of Physics, 85 Lomb Memorial Drive,
Rochester Institute of Technology, Rochester, NY 14623}

\begin{abstract}
In galactic nuclei with sufficiently short relaxation times,
binary supermassive black holes can
evolve beyond their stalling radii via continued interaction
with stars.
We study this ``collisional'' evolutionary regime using both fully 
self-consistent $N$-body integrations and approximate Fokker-Planck models.
The $N$-body integrations employ particle numbers up to $0.26\times 10^6$
and a direct-summation potential solver;
close interactions involving the binary are treated using
a new implementation of the Mikkola-Aarseth chain regularization algorithm.
Even at these large values of $N$, two-body scattering occurs at 
high enough rates in the $N$-body simulations that the binary is never 
fully in the diffusively-repopulated (i.e. large-$N$) loss cone regime, 
which precludes a simple scaling of the results to real galaxies.
The Fokker-Planck model is used to bridge this gap; it includes,
for the first time in this context,
binary-induced changes in the stellar density and potential.
The Fokker-Planck model is shown to accurately reproduce the
results of the $N$-body integrations,
and is then extended to the much larger $N$ regime of real galaxies.
Analytic expressions are derived that accurately
reproduce the time dependence of the binary semi-major
axis as predicted by the Fokker-Planck model.
Gravitational radiation begins to dominate the binary's evolution
after a time that is always comparable to, or less than, the relaxation
time measured at the binary's gravitational influence radius;
the observed correlation of nuclear relaxation time with 
velocity dispersion implies that coalescence in
$\le 10$ Gyr will occur in nuclei with $\sigma \lap 80$ km s$^{-1}$,
i.e. with binary black hole mass $\lap 2\times 10^6\msun$.
The coalescence time depends only weakly on 
binary mass ratio.
Formation of a core, or ``mass deficit,'' is shown to result 
from a competition between ejection of stars by the binary and
re-supply of depleted orbits via two-body scattering.
Mass deficits as large as $\sim 4$ times the binary mass are produced
before the gravitational radiation regime is reached;
however, after the two black holes
coalesce, a Bahcall-Wolf cusp appears around the
single hole in approximately one relaxation time, resulting in
a nuclear density profile consisting of a flat
core with an inner, compact cluster, similar to what is
observed at the centers of low-luminosity elliptical galaxies.
We critically evaluate recent claims that 
binary-star interactions can induce rapid coalescence
of binary supermassive black holes 
even in the absence of loss cone refilling.

\end{abstract}

\section{Introduction}

This paper is the third in a series investigating
the evolution of binary supermassive black holes at the
centers of galaxies.
A massive binary hardens via exchange of energy and angular momentum
with passing stars, but this process is self-limiting,
since the interacting stars are ejected from the nucleus with
velocities of order the relative velocity of the two
black holes.
Continued hardening of the binary requires a repopulation
of the depleted orbits.
Paper I \citep{MM:03} discussed various mechanisms by
which this can occur, including collisional loss-cone
repopulation, secondary slingshot, chaotic stellar orbits,
and Brownian motion of the binary.
These different mechanisms typically obey different
scalings of the binary hardening rate with the 
number $N$ of stars and with time; in the large-$N$ limit and in
a spherical or axisymmetric potential, the hardening
rate (defined as the rate of change of the binary's
energy) is predicted to scale roughly as $N^{-1}$,
i.e. inversely with the relaxation time,
and hence to be very small for values of $N$ characteristic
of massive elliptical galaxies \citep{Valtonen:96,Yu:02}.

In Paper II \citep{BMS:05}, a direct-summation $N$-body
code, combined with a parallel GRAPE cluster,
was used to carry out integrations of binary evolution
in galaxy models with large, low-density cores.
Because of their low central density, the relaxation
time at the center of these models was relatively
long (compared with orbital periods),
and collisional loss cone
refilling was shown to occur at a lower rate than the
loss of stars to the binary, i.e. the binary's loss
cone remained nearly empty.
This is the same (``diffusive'') regime believed to characterize binary
evolution in real galaxies \citep{MM:01}.
The $N$-body hardening rates were compared with the predictions
of simple loss-cone theory and found to be in reasonable agreement.

The Plummer models used in Paper II were not good representations
of real galaxies.
In this paper, we present a new set of simulations based on
galaxy models that more closely approximate real galaxies,
with power-law central density cusps.
In order to deal efficiently with interactions
involving the binary, we incorporate 
the Mikkola-Aarseth chain-regularization algorithm 
\citep{MA:90,MA:93} into our $N$-body code, including 
both the effects of nearby stars as perturbers of the chain,
and the effects of the chain on the surrounding stars.
The resulting $N$-body algorithm is coupled with a
GRAPE-6 special-purpose computer and used to carry out
extended integrations of binaries with various 
values of $N$, up to the limit $N\approx 0.26\times 10^6$
set by the GRAPE's memory.
In order to more accurately characterize the $N$-dependence
of the evolution, multiple integrations are carried out
starting from different random realizations of the same initial
conditions and averaged.

Even at the large values of $N$ allowed by the GRAPE-6, 
two-body (star-star)
scattering occurs at a high enough rate in the $N$-body simulations
that the binary is never fully in the empty-loss-cone regime.
This fact precludes a simple scaling of the $N$-body results
to real galaxies.
We therefore develop a Fokker-Planck model that can be
applied to nuclei with any value of $N$, i.e. any value 
of $\mtwo/m_\star$,
where $\mtwo\equiv M_1+M_2$ and $m_\star$ are the mass of the binary and
of a single star, respectively.
Our Fokker-Planck model is unique in that it allows for the joint 
evolution of the binary and of the stellar nucleus; it can
therefore reproduce the creation of a core, or ``mass deficit''
\citep{MM:02}, as the binary ejects stars.
The Fokker-Planck model is first tested by comparison
with the $N$-body results, and is then applied to the much 
larger-$N$ regime of real galaxies.
In this way we are able to make the first
detailed predictions about the joint evolution of massive
binaries and stars at the centers of galaxies.

The time scale that limits binary evolution in our models
is the relaxation time, defined as 
the time for (mostly distant)
gravitational encounters between stars to establish a locally 
Maxwellian velocity distribution.
Assuming a homogenous and isotropic distribution of equal-mass stars,
the relaxation time is approximately
\begin{subequations}
\begin{eqnarray}
\tr &\approx& {0.34\sigma^3\over G^2 \rho m_\star \ln\Lambda} \\
&\approx&
1.2\times 10^{10}\ {\rm yr}\ \sigma_{100}^3\ \rho_5^{-1} 
\tilde{m}_\star^{-1} \ln\Lambda_{15}^{-1}
\label{eq:tr}
\end{eqnarray}
\end{subequations}
\citep{Spitzer:87}.
Here, $\sigma_{100}$ is the 1d stellar velocity dispersion
in units of $100$ km s$^{-1}$, 
$\rho_5$ is the stellar mass density in units of $10^5 M_\odot\ {\rm pc}^{-3}$,
$\tilde{m}_\star=m_\star/M_\odot$, 
and $\ln\Lambda_{15}=\ln\Lambda/15$, 
where $\ln\Lambda$ is the Coulomb logarithm and
$\Lambda\approx 0.4N$ \citep{Spitzer:87}.

\begin{figure}
\centering
\includegraphics[scale=0.4]{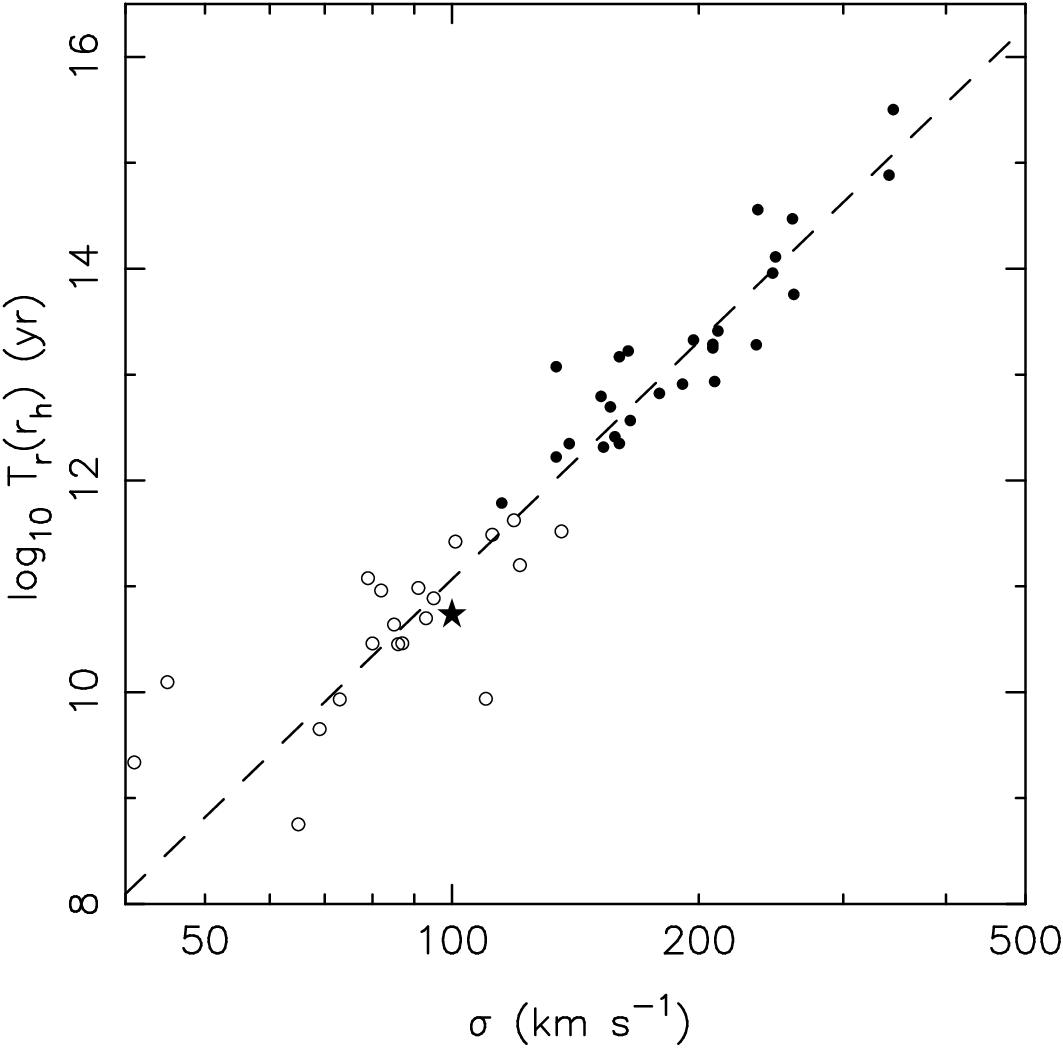}
\caption{Relaxation times, measured at the supermassive
black hole's influence radius, 
in the ACS/Virgo sample of galaxies \citep{ACS1},
versus the central stellar velocity dispersion.
Filled symbols are galaxies in which the black hole's
influence radius is resolved; star is the Milky Way.}
\label{fig:trrh}
\end{figure}

Figure~\ref{fig:trrh} shows estimates of $\tr$,
measured at the supermassive black hole's influence
radius $r_h$, for the ACS/Virgo sample of early-type galaxies \citep{ACS1}.
The influence radius was defined in the usual way via
\beq
M_\star(r_h) = 2\mh
\eeq
and the black hole mass was inferred from the measured value of
$\sigma$ via the the $\mh-\sigma$ relation,
\beq
\mh\approx 5.72\times 10^6 M_\odot \sigma_{100}^{4.86}
\label{eq:msigma}
\eeq
\citep{FF:05}.
A stellar mass of $1M_\odot$ was assumed.

Figure~\ref{fig:trrh} reveals a tight correlation between
$\tr(r_h)$ and $\sigma$.
A least-squares fit to the points (shown as the dashed line
in the figure) gives
\begin{subequations}
\begin{eqnarray}
\tr(r_h) &\approx& 1.16\times 10^{11}\ {\rm yr}\ \sigma_{100}^{7.47}\\
         &\approx& 8.0\times 10^9\ {\rm yr}\ M_{\bullet,6}^{1.54}
\end{eqnarray}
\label{eq:trsigma}
\end{subequations}
where $M_{\bullet,6}\equiv\mh/10^6M_\odot$.
The  results presented in this paper are only relevant
to galaxies in which the nuclear relaxation time is
not much longer than galaxy lifetimes; 
according to Figure~\ref{fig:trrh}, this is the case
for galaxies with $\sigma\lap 80$ km s$^{-1}$.
This is roughly the velocity dispersion near the
center of the Milky Way;
hence, the sort of evolution that is modelled here is
most relevant to spheroids that are not much brighter
than the Milky Way bulge.

The $N$-body techniques are described in \S2 and \S3 and
the results  of the $N$-body integrations are presented
in \S4.
In \S5 the Fokker-Planck model is described and compared
with  the $N$-body results.
Predictions of the Fokker-Planck model in the large-$N$
regime corresponding to real galaxies are presented in \S6.
\S7 and \S8 discuss the implications for evolution of binary
supermassive black holes in real galaxies,
and \S9 presents a critical comparison
with other proposed models of binary evolution.
\S10 sums up.

\section{$N$-Body Techniques}

Our $N$-body algorithm was an adaptation of the ${\rm NBODY1}$
code of \cite{Aarseth:99} to the GRAPE-6 special purpose hardware.
The code uses a fourth-order Hermite integration scheme
with individual, adaptive, block time steps \citep{Aarseth:03}.
For the majority of the particles, the forces and force
derivatives were calculated via a direct-summation scheme using
the GRAPE-6.
More details of the particle advancement scheme can be found in Paper II.
As discussed there, the code contains two parameters that affect the
speed and accuracy of the calculation,
the particle softening length $\epsilon$
and the time-step accuracy parameter $\eta$.

Close encounters between the massive particles
(``black holes''), or between black holes and stars,
require prohibitively small time steps in such a scheme.
To avoid this situation, we adopted a chain regularization
algorithm for the critical interactions \citep{MA:90,MA:93},
as follows.
Let ${\bf r}_i$, $i=1,...,N$ be the position vectors of the particles.
We first identify the subset of $n$ particles to be included
in the chain; the precise criterion for inclusion is
presented below,
but in the late stages of evolution, the chain always
included the two black holes as its lowest members.
We then search for the particle that is closest to either end
of the chain and add it; this operation is repeated recursively
until all $n$ particles are included.
Define the separation vectors
${\bf R}_i = {\bf r}_{i+1} - {\bf r}_i$
where ${\bf r}_{i+1}$ and ${\bf r}_i$ are the coordinates
of the two particles making up the $i$th link of the chain.
The canonical momenta ${\bf W}_i$ corresponding to the
coordinates ${\bf R}_i$ are given in terms of the old momenta
via the generating function
\beq
	S = \sum_{i=1}^{n-1} {\bf W}_i\cdot ({\bf r}_{i+1} - {\bf r}_i).
\eeq
Next, we apply KS regularization \citep{KS:65} to the chain vectors,
regularizing only the interactions between neighboring
particles in the chain.
Let ${\bf Q}_i$ and ${\bf P}_i$ be the KS transformed ${\bf R}_i$ and
${\bf W}_i$ coordinates.
After applying the time transormation $\delta t= g \delta s$,
$g = 1/L$, where $L$ is the Lagrangian of the system
($L = T - U$, where $T$ is the kinetic and $U$ is the
potential energy of the system).
 We obtain the regularized Hamiltonian
$\Gamma = g (H({\bf Q}_i,{\bf P}_i) - E_0)$,
where $E_0$ is the total energy of the system.
The equations of motion are then
\beq
{\bf P}_i' =
- {\partial \Gamma\over \partial {\bf Q}_i} \;,
\quad
{\bf Q}_i' = {\partial \Gamma\over \partial {\bf P}_i}
\eeq
where primes denote differentiation with respect to the time
coordinate $s$.
Because of the use of regularized coordinates, these
equations do not suffer from singularities,
as long as care is taken in the construction of the chain.

Since it is impractical to include all $N$ particles in the
chain, we must consider the effects of external
forces on the chain members.
Let ${\bf F}_j$ be the perturbing acceleration acting on
the $j$th body of mass $m_j$.
The perturbed system can be written in Hamiltonian form by
simply adding the perturbing potential:
\beq
\delta U = -\sum_{j=1}^n m_j {\bf r}_j \cdot {{\bf F}_j}(t).
\eeq

Only one chain was defined at any given time.
At the start of the $N$-body integrations,
there was no regularization, and all particles were
advanced using the variable-time-step Hermite scheme.
The first condition that needed to be met before
``turning on'' the chain was that one of 
the particles (including possibly a black hole)
achieved a time step shorter than $t_{chmin}$ and
passed a distance from one of the black holes smaller
than $r_{chmin}$.
If this condition was satisfied, it was then checked whether
the encounter resulted in a deflection angle greater than
$2\delta=\pi/2$, where
\beq
	\cos \delta = \left[ 1 + \frac{R^2 V_0^4}{G^2(m_1+m_2)^2} \right]^{-1/2} \; ;
\eeq
here $R$ is the impact parameter, $V_0$ is the pre-encounter
relative velocity, and $m_1$ and $m_2$ are the masses of the 
two particles.
This condition is equivalent to
\beq
	m_1 + m_2 > R V_0^2.
\eeq
Each star closer to the black hole
than $r_{chmin}$ was then added to the chain,
and the two black holes were always included.
The values of $t_{chmin}$ and $r_{chmin}$
were determined by carrying out test runs; we adopted
$t_{chmin} \approx 10^{-5} - 10^{-6}$ and $r_{chmin} \approx 10^{-4} - 10^{-3}$
in standard $N$-body units.

\begin{figure}
\centering
\includegraphics[scale=0.55,angle=0.]{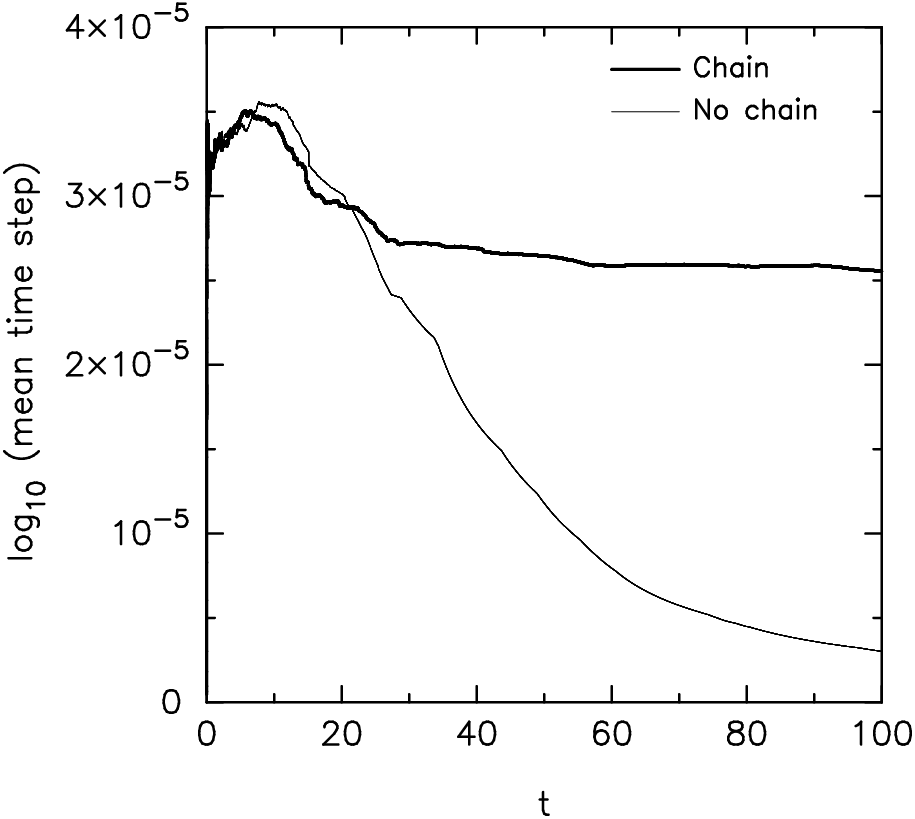}
\caption{The average time step, as defined in the text, 
during two integrations of a binary 
black hole at the center of a Dehen-model galaxy.
$N=20,000$, and the softening length and time-step parameters of the $N$-body
code were $\epsilon=10^{-6}, \eta=0.01$.
In the absence of the chain, the average time step drops to very
low values once the binary begins to harden. }
\label{fig:tstepav}
\end{figure}

\begin{figure}
\centering
\includegraphics[scale=0.6,angle=0.]{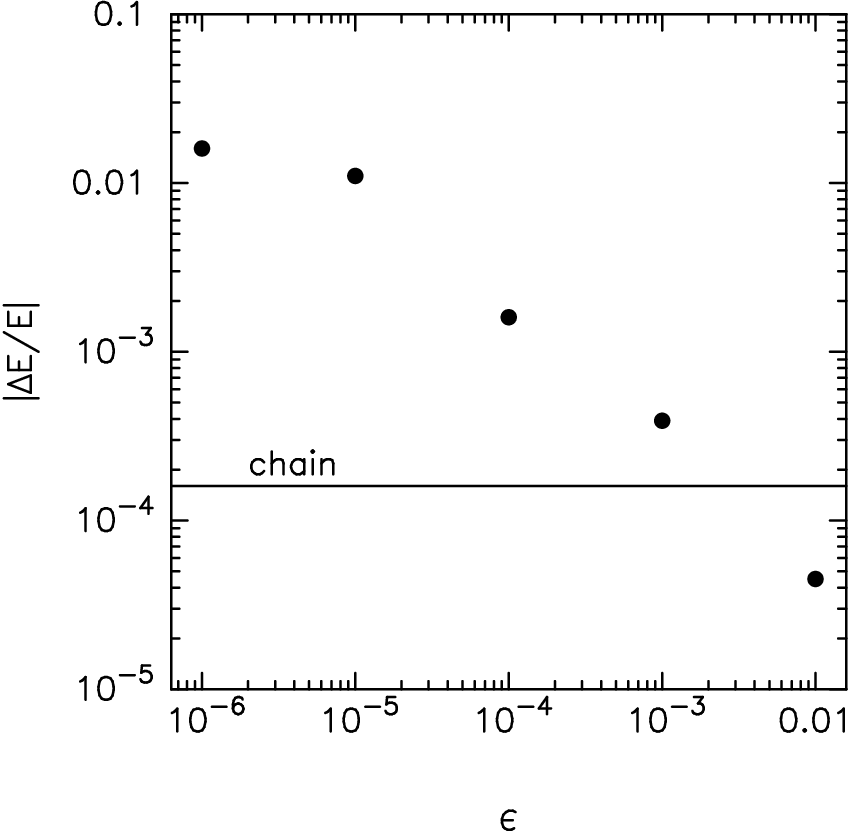}
\caption{Relative energy error over $100$ time units of 
of a set of integrations like those in Fig.~\ref{fig:tstepav}, for
various values of the softening length $\epsilon$, and with
the chain.}
\label{fig:relenerror}
\end{figure}

The chain's center of mass was a pseudoparticle as seen by the
$N$-body code and was advanced by the Hermite scheme in the same 
way as an ordinary particle. 
However, when integrating the trajectories of stars near to the chain,
it is essential to resolve the inner structure of the chain.
Thus for stars inside a critical $r_{crit1}$ radius around the
chain, the forces from the individual chain members were taken 
into account. 
The value of $r_{crit1}$ was set by the size of the chain to be
$r_{crit1} = \lambda R_{ch}$ with $R_{ch}$ the spatial size 
of the chain and $\lambda = 100$.
In addition, the equations of motion of the chain particles 
must include the forces exerted by a set of external perturber stars. 
Whether or not a given star was listed as a perturber was determined by
a tidal criterion: 
$r < R_{crit2} = (m/m_{chain})^{1/3} \gamma_{min}^{-1/3} R_{ch}$
where $m_{chain}$ represents the mass of the chain, 
$m$ is the mass of the star, and $\gamma_{min}$ was chosen to be $10^{-6}$;
 thus $r_{crit2} \approx 10^{2} (m/m_{chain})^{1/3}  R_{ch}$.

The membership of the chain changed under the evolution of the system. 
Stars were captured into the chain if their orbits approached the binary 
closer than $R_{ch}$. Stars were emitted from the
chain if they got further from both of the black holes than 
$1.5 R_{ch}$. The difference between the emission
and absorption distances was chosen to avoid a too-frequent 
variation of the chain membership.
When the last particle left the chain,
the chain was eliminated and the integration turned back
to the Hermite scheme, until a new chain was created.

In what follows, we refer to the $N$-body code without chain
as NB1, and the code including chain as CHNB1.
We carried out a number of tests to see how the performance 
and accuracy of the NB1 code were affected by inclusion of the chain.
Typically, one integration step of the chain required about five 
times as much cpu time as a single call to the GRAPE-6, due to
the complex nature of the chain and the generally large number
of perturber particles. 
Thus our code is quicker than a basic Hermite scheme code (NB1) 
only if the smallest time steps are about an order of magnitude
shorter than the next smallest time steps,
and if in addition those particles would be assigned to the chain. 
It is easy to show that in the case of a galaxy including a central,
massive binary system this condition is usually fulfilled. 
The Hermite time step of the binary is considerably smaller than 
the time steps of the stars, due to their close orbit and fast evolution.
Of course, the performance of both codes depends on
the two parameters $\eta$ (time step parameter) and $\epsilon$
(particle softening length) that determine the accuracy of the
star-star interactions.
In what follows, we fixed $\eta=0.01$ based on the results
of the tests in Paper II.
In CHNB1, $\epsilon$ was always set to zero.

\begin{figure}
\centering
\includegraphics[scale=0.8]{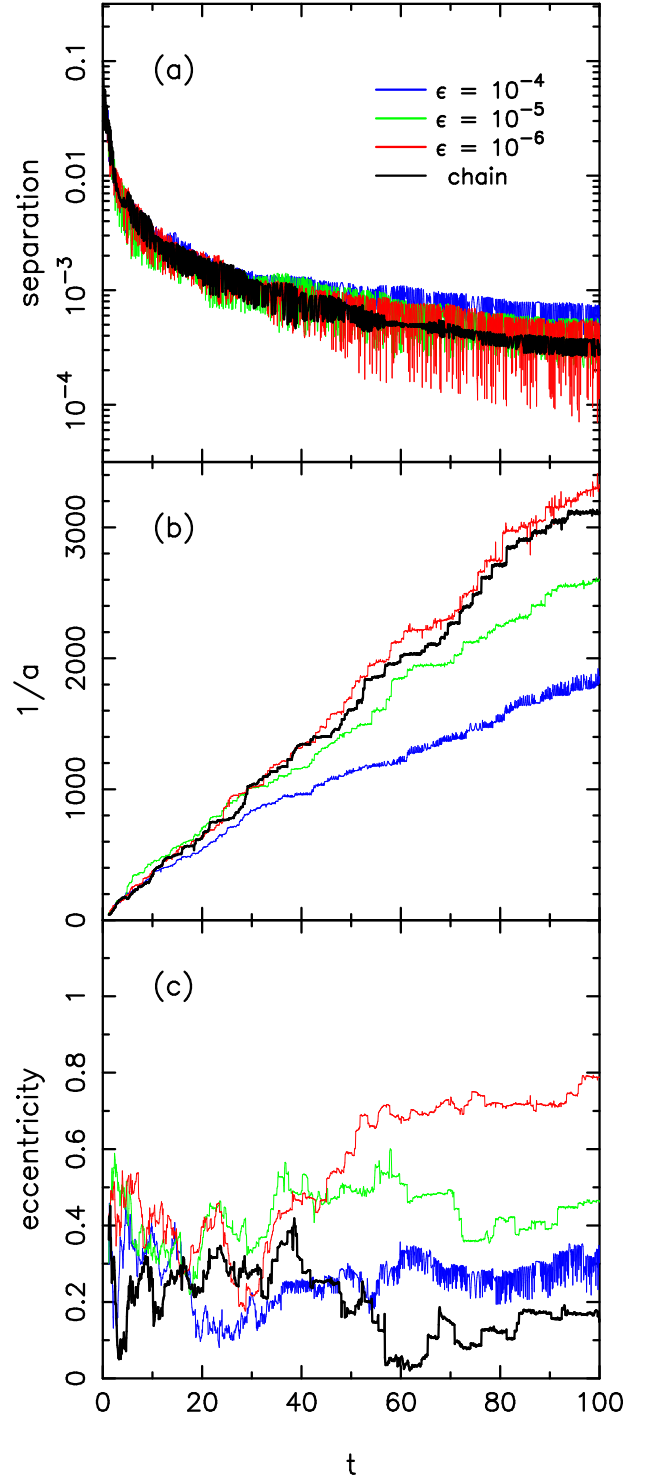}
\caption{Results of a set of test integrations with and without
the chain. Initial conditions consisted of a binary of mass
$M_1=M_2=0.005$ and separation $0.1$, in a Dehnen-model
galaxy with $\gamma=1.2$ and $N=20,000$.}
\label{fig:testint}
\end{figure}

\begin{figure*}
\centering
\includegraphics[scale=0.85]{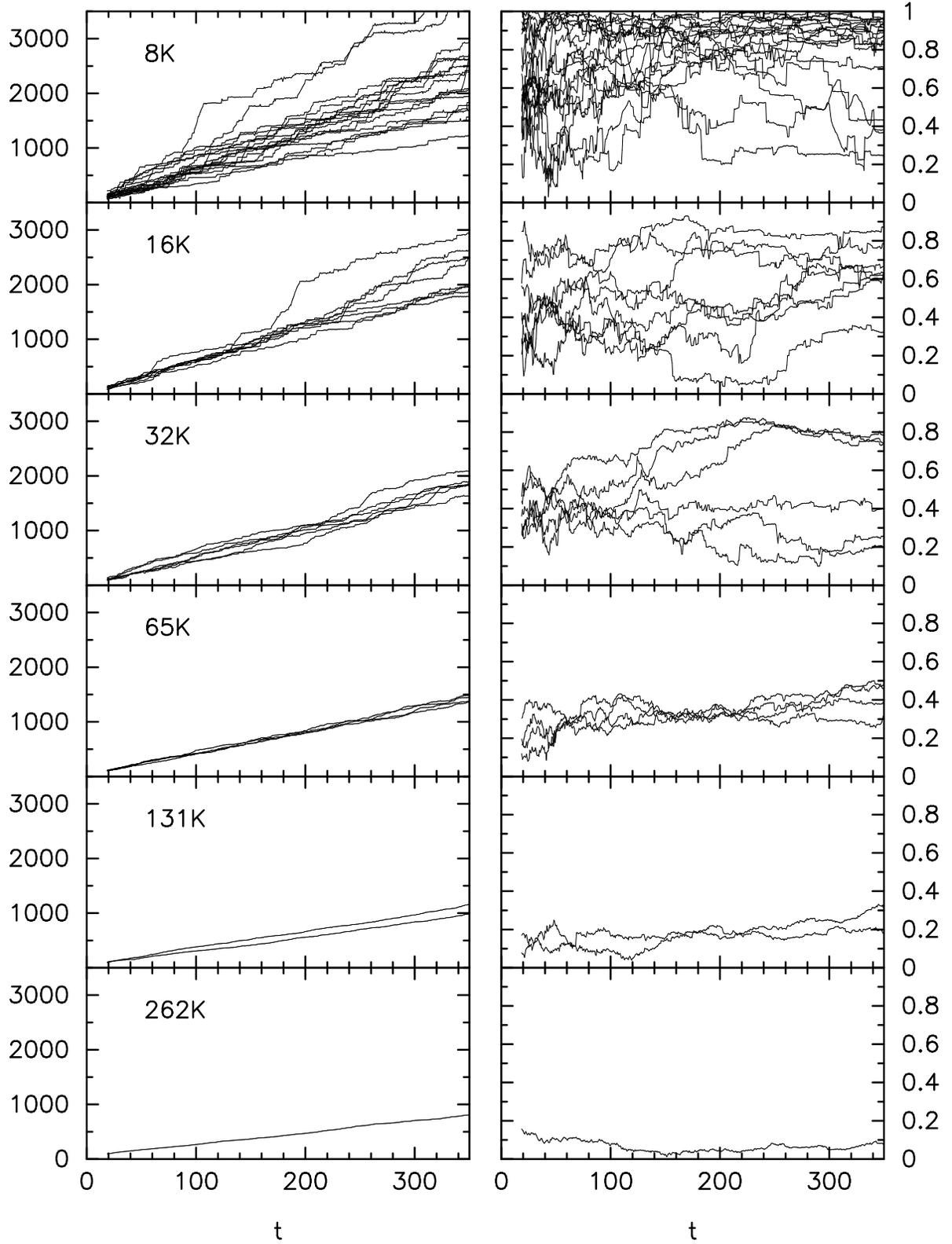}
\caption{Evolution of $1/a$, the inverse binary semi-major
axis (left column), and $e$, the binary eccentricity
(right column), in the full set of $N$-body integrations.
\label{fig:all}}
\end{figure*}

Figures~\ref{fig:tstepav}-\ref{fig:testint}
show the results of our performance tests.
Figure~\ref{fig:tstepav} plots the average time step as a function of time
in both codes, for integrations of a binary black hole in a
galaxy model following Dehnen's (1993) density law:
\begin{equation}
\rho (r) = {(3 - \gamma)M_{gal}\over 4 \pi} {a \over r^{\gamma}(r + a)^{4 - \gamma}}
\label{eq:dehnen}
\end{equation}
with $\gamma=1.2$ and $N=20,000$ particles.
The two black holes had equal masses,
$M_1=M_1=0.01M_{gal}$, and were placed intially on a circular orbit
with separation $0.10a$.
We defined the average time step as 
$t/N_{time steps}(t)$, where $N_{time steps}(t)$ was the total number of
integration time steps until time $t$, including only the time
steps of particles outside the chain.
It can be seen that in the early stages of the evolution, 
the time steps are about the same in both cases.
However as the binary becomes harder, the NB1 time steps become 
smaller and smaller,
in order to achieve the necessary precision in the integration 
of the binary.
In the code with the chain, the average time step hardly changes
after the binary begins to harden. 
The binary is integrated
by the regularized equations, hence the step size of the NB1 integration
remains relatively large. 
The average step size of the CHNB1 code was about $2.5 \times 10^{-5}$
(in units where $G=M_{gal}=a=1$),
while in the NB1 integration it was $3 \times 10^{-6}$.
The resulting net speed-up with the chain is more than a factor of two.

Figure~\ref{fig:relenerror} shows the energy error in a set of integrations 
of the same model but with various different softening lengths 
$\epsilon$, compared with the energy error in an integration
with the chain (and $\epsilon=0$).
It can be seen that the energy conservation of the CHNB1
code is about as good as the best results from the NB1 integrations.
However, the latter occur when the softening length is very
large, much too large for accurate integration of the binary.
This is shown in Figure~\ref{fig:testint}.
It is evident from that figure that with larger softening lengths, 
$\epsilon \gap 10^{-5}$, the integration of the binary is not 
very accurate.
However, even with very small $\epsilon$, the
evolution of the distance between the black holes includes 
"spiky" behaviors, due apparently to 
the very sensitive nature of the eccentricity 
evolution with respect to the precision of the integration
(Figure~\ref{fig:testint}b,c).

These results suggest that chain regularization
is an accurate and efficient way to integrate
binary black holes at the centers of galaxies.
It can keep track of the evolution of the binary with 
high precision, and the calculation time is substantially 
faster than a plain Hermite integration when the
latter is used with a reasonable (i.e. sufficiently
small) softening parameter.

\section{Initial Conditions}

All of our integrations adopted Dehnen's
model, equation~(\ref{eq:dehnen}), for the initial galaxy,
with $\gamma=0.5$.
To this model were added two particles,
the ``black  holes,'' with masses $M_1=M_2=0.005$ in
units of the galaxy mass.
(Henceforth we write $\mtwo\equiv M_1+M_2$.)
The black holes were placed symmetrically about  the center
of the galaxy at $x=\pm 0.1$.
The initial velocities of the black holes were chosen to
be $v_y=\pm 0.16$ yielding nearly circular initial orbits.
These initial conditions are similar to those adopted in
some earlier studies (\cite{Quinlan:97,Nakano:99}) although
they are probably less realistic than initial conditions 
that place one of the two massive particles exactly at the center
(e.g. \cite{MS:06}).
Henceforth we adopt units such that the gravitational
constant $G$, the total mass in stars $M_{gal}$, 
and the Dehnen scale length $a$ are equal to one.
In these units, the crossing time
$(GM_{gal}/a^3)^{-1/2}$ is also equal to one.

A standard expression for $r_h$, the radius of influence of
a single black hole at the center of a galaxy, is
\beq
M_\star(r_h)=2\mh
\eeq
with $\mh$ the black hole mass and $M_\star(r)$ the
mass in stars within a sphere of radius $r$ .
The semi-major axis length of a ``hard'' binary is 
sometimes defined in terms of $r_h$ as
(e.g. \cite{MW:05}) 
\beq
a_h = {\q\over \left(1+\q\right)^2} {r_h\over 4}
\label{eq:ah}
\eeq
with $\q\equiv M_2/M_1\le 1$ the binary mass ratio,
and we adopt that definition here.
Setting $\mh=\mtwo = 0.005+0.005 = 0.01$ and $\q=1$,
the values of $r_h$ and $a_h$ for our $N$-body models
are 
\beq
r_h = 0.264,\ \ \ \ a_h=0.0165.
\label{eq:rhah}
\eeq
These expressions ignore the changes that the two black holes
induce in the mass distribution of the galaxy when
forming a hard binary, but are useful as points
of reference.

Based on the results of Papers I and II, once the binary
has interacted with and ejected most of the stars on
intersecting orbits, its subsequent evolution is dependent
on the continued scattering of stars into its sphere
of influence; since the scattering time scale increases
with $N$, the binary's decay rate should 
decrease as $N$ increases.
In order to better characterize this $N$-dependence,
the initial conditions were realized using six different 
values of $N$,
$N = (8192$, $16384$, $32768$, $65536$, $131072$, $262144$),
or $N=2^p$, $p=(13,14,15,16,17,18)$. 
(In what follows, we refer to these different $N$-values
via the shorthand 8K, 16K, ..., 262K).
The largest of these $N$ values is close to
the maximum number of particles that can be handled in the GRAPE-6
memory.
In order to decrease the ``noise'' associated with
the evolution for small $N$, 
we carried out $n_{int}$ multiple integrations at each $N$,
in which the initial stellar postions and velocities
were calculated using different seeds for the random number generator.
All of these integrations were continued until a time
$T_{max}=350$;
when scaled to a typical luminous elliptical galaxy
with crossing time $\sim 10^8$ yr,
this corresponds to $\sim 10^{10}$ yr.
Table 1 gives the parameters of the $N$-body integrations.

\begin{table}
\begin{center}
\caption{Parameters of the $N$-body integrations \label{tbl-2}}
\begin{tabular}{ccc}
\tableline\tableline
Name & $N$ & $n_{int}$ \\
\tableline
8K & 8192 & $18$ \\ 
16K & $16384$ & $8$ \\
32K & $32768$ & $6$ \\
65K & $65536$ & $4$ \\
131K & $131072$ & $2$ \\
262K & $262140$ & $1$ \\
\tableline
\end{tabular}
\end{center}
\end{table}

\section{$N$-Body Results}

Initially the two black holes move on nearly independent orbits 
about the center of the galaxy.
The orbits decay, and at $t\approx 10$ the black holes 
form a hard binary.
After this, the semimajor axis $a$ of the binary
shrinks as the two black holes interact
with stars and eject them from the nucleus via the
gravitational slingshot.
Figure~\ref{fig:all} shows the  evolution of $1/a$
and $e$, the orbital eccentricity, 
in the full set of integrations for $t\ge 20$.
The scatter in the values of $1/a$ and $e$ at a given
time is considerable in the integrations with smallest $N$.
Nevertheless a clear trend is apparent:
both $1/a$ and $e$ evolve less, on average,
as $N$ is increased.

\subsection{Binary Hardening}

\begin{figure*}
\centering
\includegraphics[scale=0.8,angle=-90.]{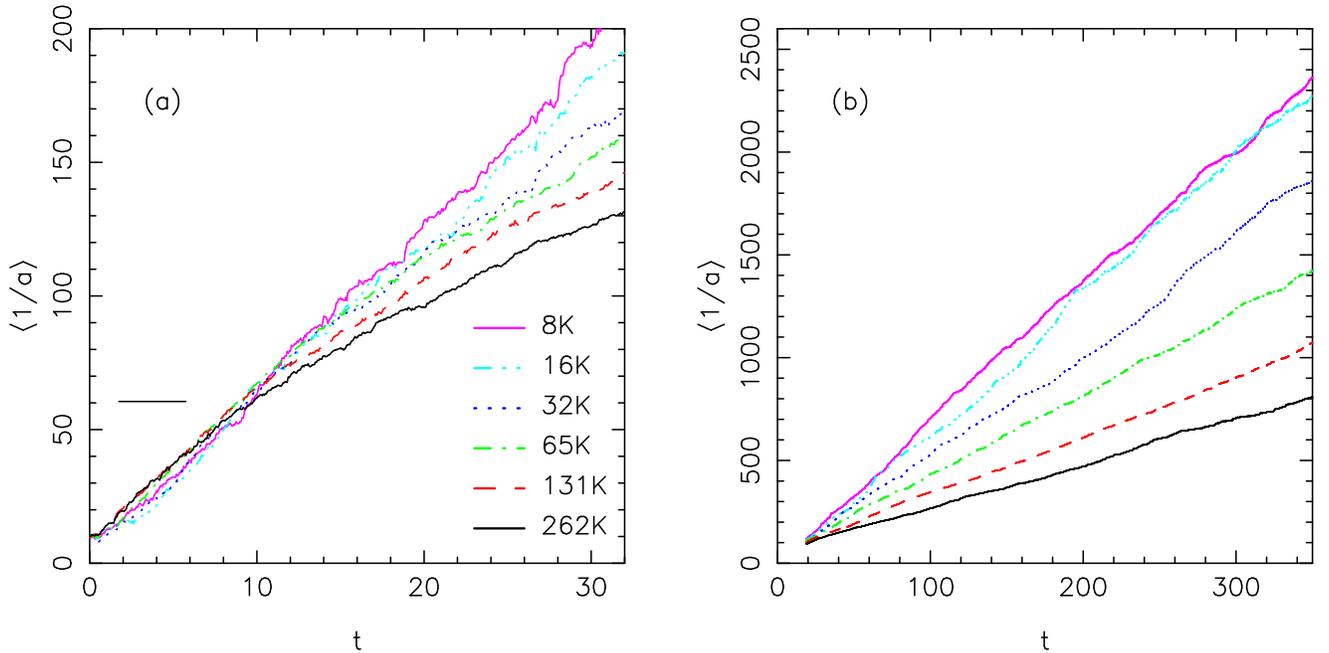}
\caption{Short-term (a) and long-term (b) evolution of the mean value of $1/a$
in the $N$-body integrations.
Horizontal line in panel (a) indicates approximately where the 
transition occurs between $N$-independent and $N$-dependent
evolution; this is also roughly the ``stalling radius'' defined
in Merritt (2006), and the ``hard binary'' separation defined
in Yu (2002).
\label{fig:av_ainv}}
\end{figure*}

In order to clarify the $N$-dependence of the evolution,
we computed averages over the $n_{int}$ independent integrations
of $a^{-1}(t)$ and $e(t)$.
Figure~\ref{fig:av_ainv} shows the mean evolution of $1/a$
for the six different $N$ values.
The early evolution (Fig.~\ref{fig:av_ainv}a), until
$t\approx 10$, is essentially $N$-independent.
In this regime, the hardening of the binary is driven
by dynamical friction against the stars, and the
rate of binding energy increase is a function only
of the stellar density, which is the same for each of 
the $N$-body models.

At $t\approx 10$, the binary hardening rate begins to show a clear
$N$-dependence, in the sense of more gradual hardening for larger $N$.
In Merritt (2006), the separation at which this occurs
was defined as the ``stalling radius,'' since in the limit of large $N$
the binary would stop evolving at this point.
Based on Figure~\ref{fig:av_ainv}, 
$a_{stall}^{-1}\approx a_h^{-1}\approx 60$ and $t_{stall}\approx 10$.

\begin{figure}
\centering
\includegraphics[scale=0.4]{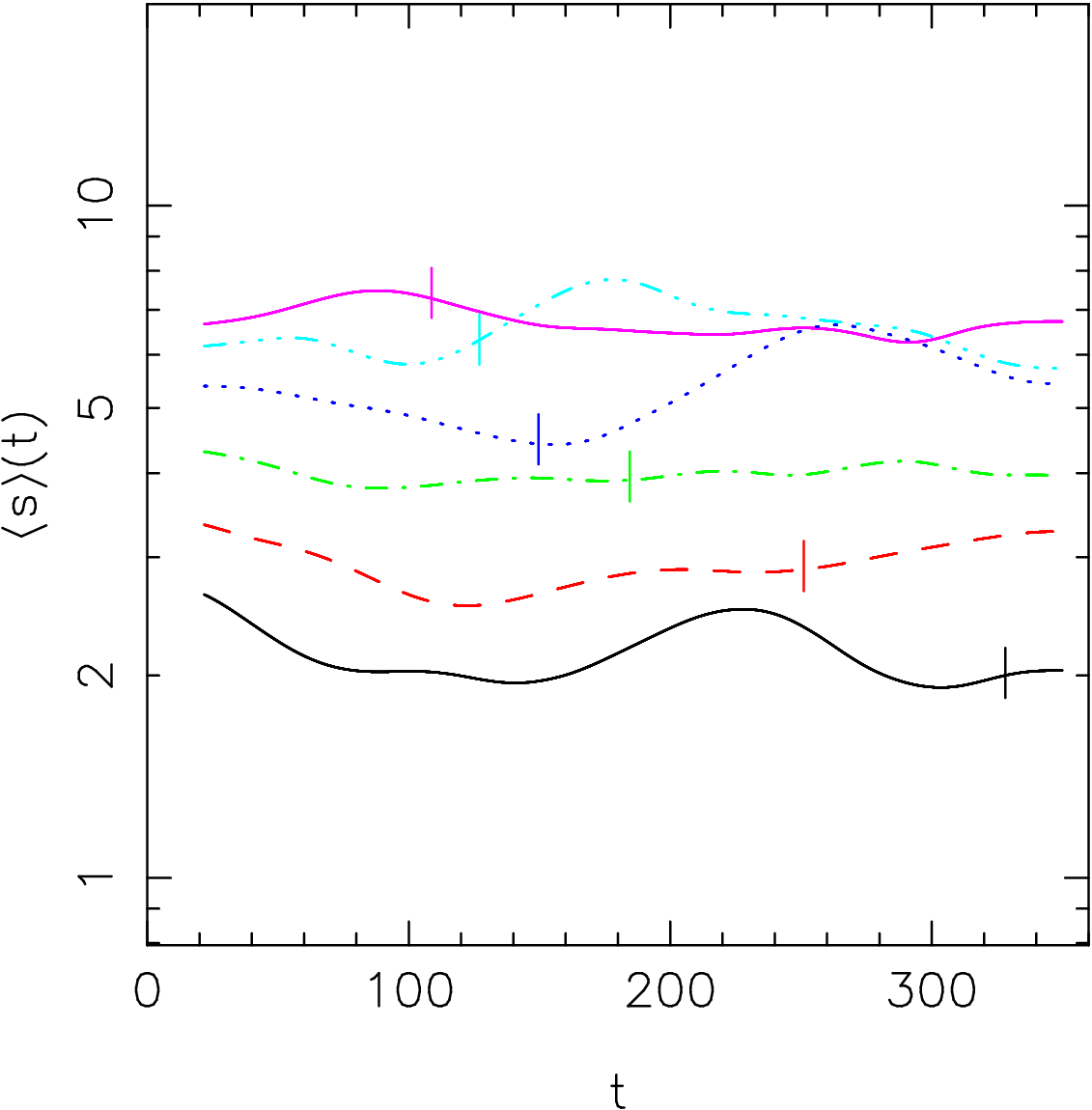}
\caption{Binary hardening rate as a function of time,
computed as an average over the $n_{int}$ independent
integrations at each $N$.
Line styles have the same meaning as in Figure~\ref{fig:av_ainv}.
Tick marks indicate where the hardening rate was 
evaluated for Fig.~\ref{fig:fpslopes}, i.e., at
$\langle a^{-1}\rangle = 750$.
\label{fig:slope}}
\end{figure}

\begin{figure}
\centering
\includegraphics[scale=0.75,angle=-90.]{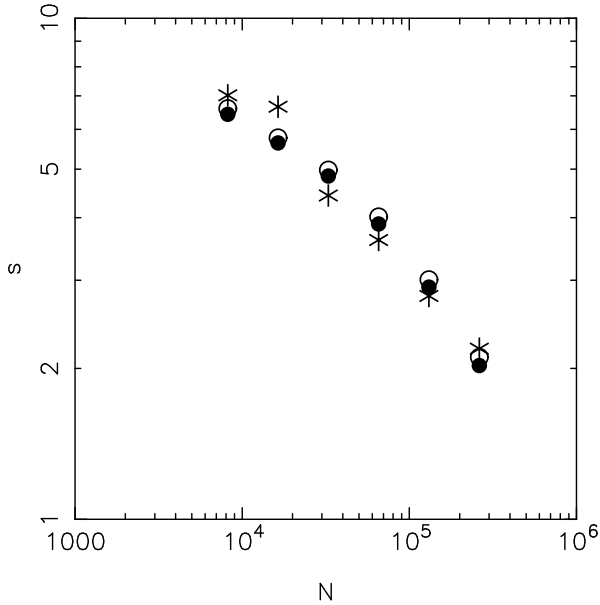}
\caption{$N$-dependence of the binary hardening rate,
computed by fitting $a^{-1}(t)$ in an interval $\Delta t = 50$
centered on $a^{-1}=700$.
{\it Asterices}: $N$-body results (Fig.~\ref{fig:av_ainv}(b)),
computed as averages over the set of ensembles at each $N$.
{\it Filled circles}: Fokker-Planck results (Fig.~\ref{fig:fp}), 
omitting the ``secondary slingshot.''
{\it Open circles}: Fokker-Planck results, including
the ``secondary slingshot.''
}
\label{fig:fpslopes}
\end{figure}

At $t\gap t_{stall}$ the $N$-dependence of the evolution is
striking (Fig.~\ref{fig:av_ainv}b).
As in Paper II, we define the instantaneous hardening rate as 
\beq
s(t)\equiv {d\over dt}\left({1\over a}\right).
\eeq
Figure~\ref{fig:slope} shows $\langle s\rangle(t)$ computed
by fitting smoothing  splines to the averaged $a^{-1}(t)$
curves of Figure~\ref{fig:av_ainv}b.
Mean hardening rates are roughly constant with time for each $N$.
The $N$-dependence of the hardening rate is shown in Figure
\ref{fig:fpslopes}.
Here, $\langle s\rangle$ was computed by fitting a straight
line to $\langle a^{-1}\rangle(t)$ in an interval
$\Delta t=50$ centered on $\langle a^{-1}\rangle(r)=750$;
in this way, the different hardening rates are being compared
at similar values of the binary semi-major axis, 
chosen to be roughly
the minimum value reached in the integration with largest $N$.
The dependence of $\langle s\rangle$ on $N$ is approximately
a power law,
\beq
\log_{10}\overline{s} \approx 2.27 - 0.357\log_{10} N .
\label{eq:sofN}
\eeq
The $\sim N^{-0.4}$ dependence is considerably flatter
than the $\sim N^{-1}$ dependence expected in a diffusively-refilled
loss cone \citep{MM:03}.
This fact precludes any simple extrapolation of the data in
Figure~\ref{fig:fpslopes} to the much larger $N$ regime
of real galaxies.

We can compare these hardening rates with the predictions
of scattering experiments in a fixed, infinite,
homogeneous background:
\beq
s\equiv {d\over dt}\left({1\over a}\right) = H {G\rho\over\sigma}
\label{eq:simple}
\eeq
with $\rho$ and $\sigma$ the mass density and 1d velocity 
dispersion of the stars,
and $H$ a dimensionless rate coefficient
that depends on the binary separation, mass ratio
and eccentricity.
For a hard, equal-mass, circular-orbit binary,
$H\approx 16$
\citep{Hills:83,Mikkola:92,Quinlan:96a,Merritt:01}.
Unfortunately, neither $\rho$ nor $\sigma$ are well defined
for our $N$-body models:
$\rho$ is formally divergent as $r\rightarrow 0$ 
(eq. \ref{eq:dehnen}),
and $\sigma$ drops to zero at the origin 
in the absence of the central binary \citep{Dehnen:93}.
We can crudely evaluate equation (\ref{eq:simple})
by setting $\rho\approx 0.595(0.338)$ and $\sigma\approx 0.216(0.244)$,
the mean and mass-weighted, rms values within a sphere of radius $0.1(0.2)$
about  the center of the (binary-free)  Dehnen model.
The results, with $H=16$, are $s\approx 44(22)$.
These are likely to be overestimates: 
the central density of the galaxy drops as the binary 
ejects stars 
and the central velocity dispersion is increased by 
the presence of the binary.
If we decrease $\rho$ by a factor of two to account for
ejections 
and set $\sigma$ equal to the rms velocity
dispersion in the $\gamma=0.5$ Dehnen model 
containing a central, $M=0.01$ point mass,
the predicted hardening rates drop to $\sim 13(8)$.
These numbers are reasonably consistent with 
the low-$N$ hardening rates shown in in Figure \ref{fig:fpslopes},
$\overline{s}\approx 6.5$, suggesting that the binary
is approximately in the ``full loss cone'' regime at
these low values of $N$.

\subsection{Eccentricity Changes}

\begin{figure}
\centering
\includegraphics[scale=0.40]{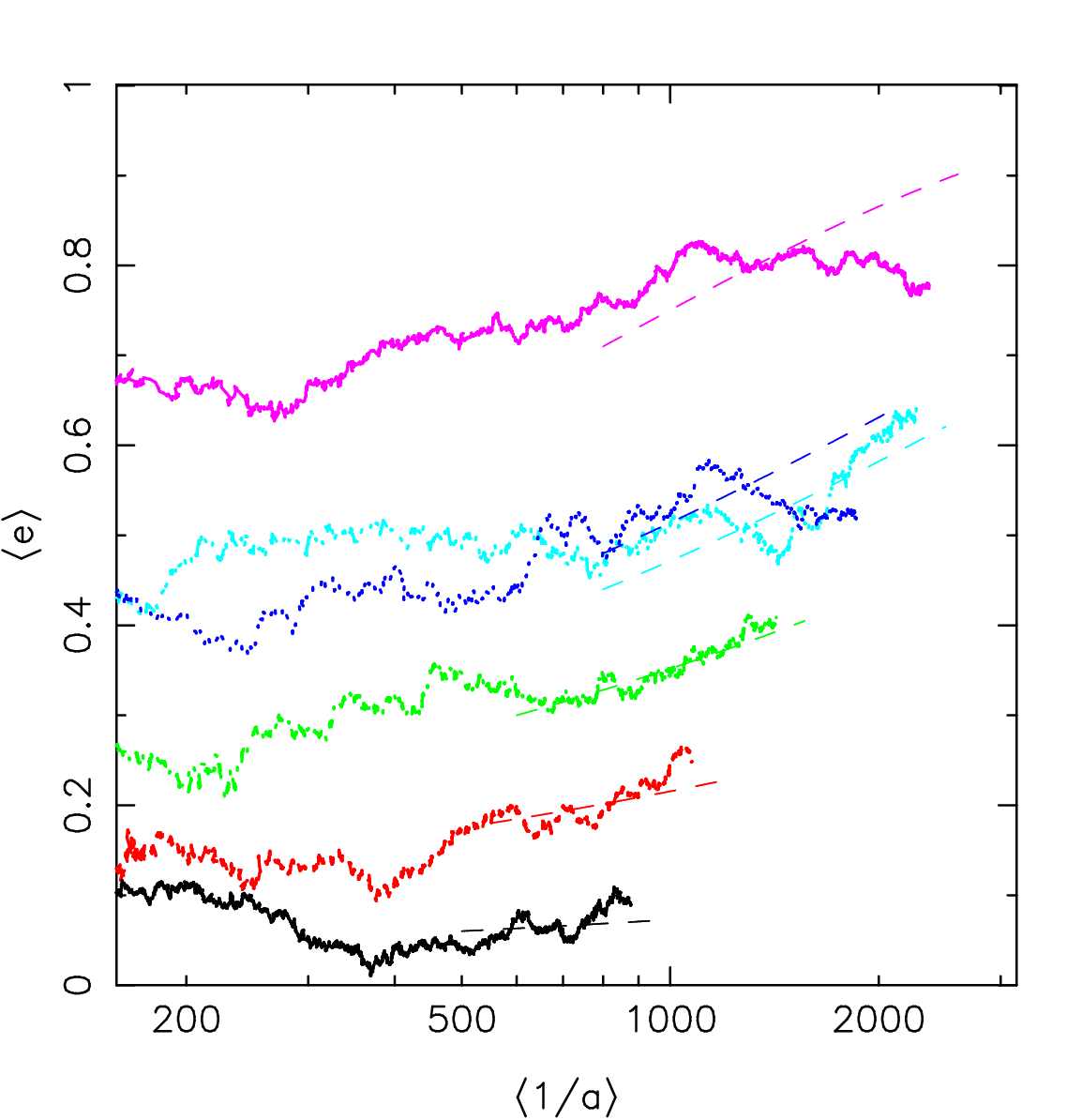}
\caption{Evolution of the mean value of $e$.
Each line is an average of the $e$ values
in the various $N$-body integrations that started
from different random realizations of the same initial
conditions.
Dashed lines show solutions to equation~(\ref{eq:anale}).}
\label{fig:av_e}
\end{figure}

The $N$-dependence of the eccentricity evolution 
(Figure~\ref{fig:av_e}) is not quite so transparent.
Although the two black holes were initially
placed on circular trajectories, perturbations
from passing stars sometimes resulted in 
very non-zero eccentricities developing around or 
even before the time the binary became hard.
This was especially true in the small-$N$ 
integrations (Fig.~\ref{fig:all});
for $N=8K$, the mass ratio between black hole and
star was only 40 and a single star-binary
interaction at early times could induce a substantial 
change in the binary's orbit.
Once established at early times, these eccentricities
tended to persist.
Spurious changes in $e$ in small-$N$-body simulations
have been noted by other authors \citep{Quinlan:97,MM:01}.
However the general trend in Figures~\ref{fig:all} and~\ref{fig:av_e} 
is clearly toward smaller eccentricities for larger $N$.

Statements about eccentricity evolution 
of massive binaries are often based
on the results of three-body scattering experiments
\citep{Mikkola:92,Quinlan:96a,Merritt:01}.
In these experiments, changes in $e$ are typically
expressed in terms of changes in $a$ as
\begin{equation}
\left\langle{de\over dt}\right\rangle = K \left\langle 
{d\over dt} \ln \left({1\over a}\right)\right\rangle
\label{eq:defK}
\end{equation}
where $K=K(e,a)$ is a dimensionless rate coefficient
and $\langle\rangle$ indicates averages over
impact parameter and velocity at infinity.
Mikkola \& Valtonen (1992) and Quinlan (1996) give
approximate analytic fits to $K_1(e,a,v_\infty)$,
the impact-parameter-averaged
rate coefficient describing changes in $e$ due
to interaction of the binary with stars of 
a {\it single} velocity $v_\infty$.
These expressions for $K_1$ can be converted into expressions for $K$ by
averaging over an assumed velocity distribution at infinity,
and Quinlan (1996, Fig.~9) shows the results of such a
calculation. 
(Sesana et al. 2006 present similar plots.)
Evolution is always found to be in the direction of increasing
eccentricity, i.e. $K\ge 0$,
excepting possibly in the case of soft, nearly-circular
binaries (Quinlan 1996, Fig.~9d-f).
Evolution rates tend to increase with increasing  hardness
of the binary, reaching maximum values of $K\approx 0.2$
for equal-mass binaries with $e\approx 0.75$ and
falling to zero at $e=0$ and $e=1$.
This is at least qualitatively consistent with 
Figure~\ref{fig:av_e},
which shows $de/d\ln(1/a)$ generally 
increasing at late times,
i.e. for larger binding energies.

Comparing these predictions quantitatively with the
$N$-body experiments is desirable, but problematic 
for a variety of reasons, the most 
important of which is probably the strong dependence
of $K$ on $\sigma/V_{bin}$, where $\sigma$ is the 
stellar velocity dispersion (assumed independent of
position) and $V_{bin}$ the binary orbital velocity.
In galaxy models like ours, $\sigma$ is a steep function
of radius near the galaxy's center
and it is not clear what value to choose.

\begin{figure}
\centering
\includegraphics[scale=0.375]{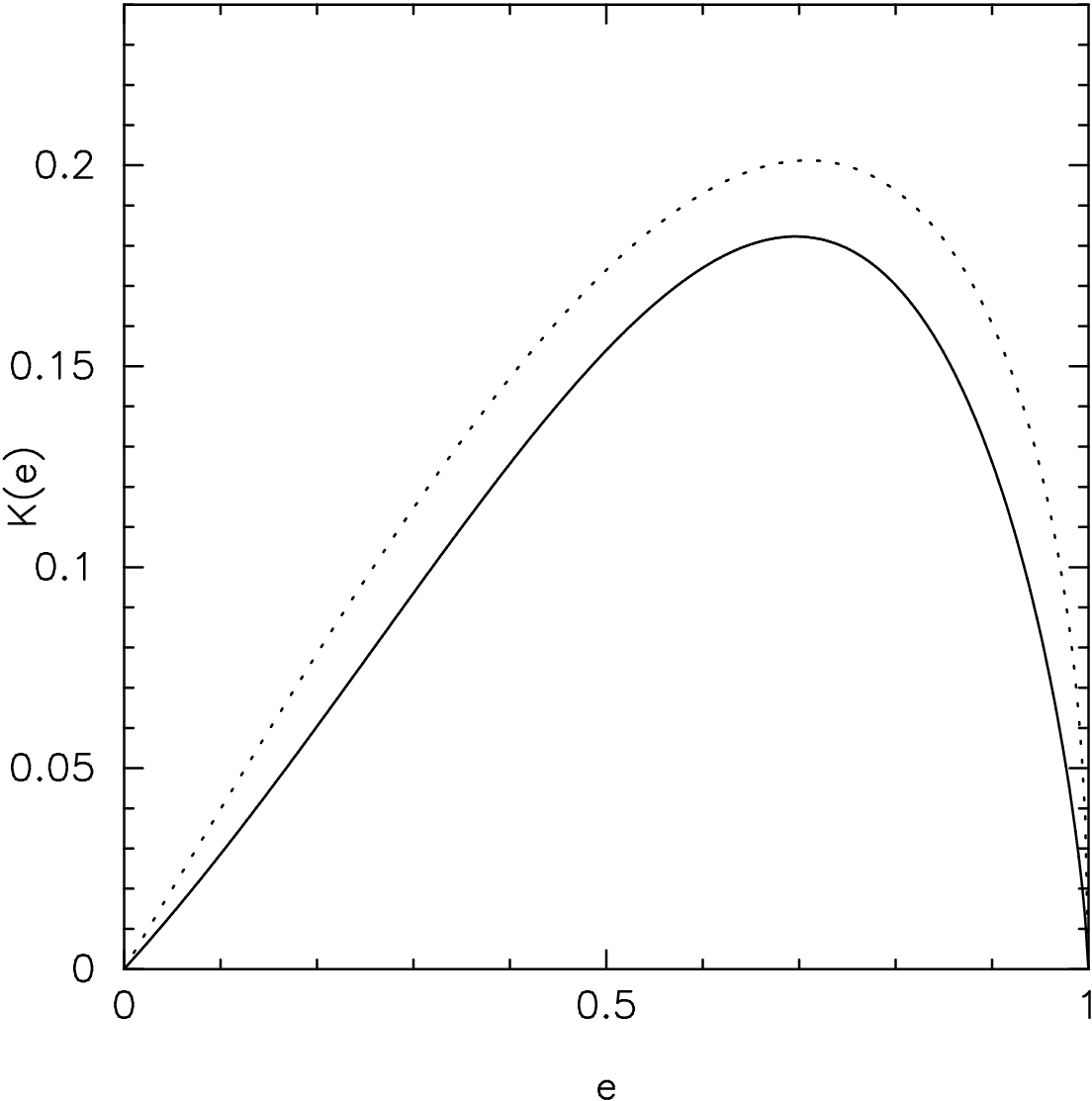}
\caption{Two approximations, derived from three-body scattering
experiments, for the coefficient $K$ (eq.~\ref{eq:defK})
describing the rate of eccentricity evolution in the limit
of large binding energy.
{\it Solid line:} Quinlan (1996); {\it dashed line:}
Mikkola \& Valtonen (1992).
The solid line was used to computed the evolutionary
tracks (dashed lines) in Fig.~\ref{fig:av_e}.
\label{fig:kofe}}
\end{figure}

In the limit of large binding energy, $V_{bin}\gg\sigma$,
the velocity at infinity is irrelevant and $K$
as determined by the scattering experiments
becomes independent of $a$.
Mikkola \& Valtonen (1992) find for $K$ in this limit
the approximate expression
\begin{subequations}
\begin{eqnarray}
K(e) &\approx& {\left(1-e^2\right)\over 2e}\left[\left(1-e^2\right)^m-1\right], \\
 m & = & 0.3e^2-0.8
\end{eqnarray}
\end{subequations}
while Quinlan (1996) gives, for an equal-mass binary in the
large-binding-energy limit,
\begin{subequations}
\begin{eqnarray}
K(e) & \approx & e\left(1-e^2\right)^{k_0}\left(k_1+k_2e\right), \\
(k_1,k_2,k_3) &=& (0.731,0.265,0.230).
\end{eqnarray}
\label{eq:Q96}
\end{subequations}
\noindent Figure~\ref{fig:kofe} shows that the two expressions
are in good agreement.

A rough value of $\sigma/V_{bin}$ in our simulations is
$\sim 2 a^{1/2}$,
where $\sigma$ has been set to $\sim 0.2$,
its mean value within a sphere of radius $0.1$
(neglecting the effects of the binary).
For $a^{-1}$ in the range $500\lap a^{-1}\lap 2500$
(Fig.~\ref{fig:av_ainv}),
this expression gives $0.1\gap \sigma/V_{bin} \gap 0.04$.
Figure~9 from Quinlan (1996) suggests that $K(e)$ 
reaches its large-binding-energy limit for
$\sigma/V_{bin}\lap 0.05$, so our simulations should be in or
near this regime at late times.

Accordingly, Figure~\ref{fig:av_e} shows solutions to
\beq
de = K(e)\ d \ln a^{-1}
\label{eq:anale}
\eeq
using Quinlan's expression for $K(e)$.
The agreement with the $N$-body results is quite reasonable,
especially for the larger values of $N$.
Nevertheless, we stress again that the final eccentricity
values  in our  $N$-body simulations are influenced
strongly by noise-induced changes in $e$ at early times,
and these changes would be much smaller in the large-$N$ regime
of real galaxies.

\subsection{Mass Deficits}

As the binary hardens, it ejects stars from the nucleus
and lowers its density.
These density changes are sometimes estimated from
scattering experiments in a fixed background like those described above, 
e.g. the change in core mass is equated with the mass 
``ejected'' by the binary.
However the fact that the binary continues to harden at late
times (Fig.~\ref{fig:av_ainv}b) implies that depopulated
orbits are continually being re-supplied.
Changes in nuclear density are therefore a competition
between ejection of stars (some of which may remain bound to the core)
and  re-population of orbits by gravitational scattering.
A number of other mechanisms can also influence the evolution
of the central density; for instance, loss of matter from the
core lowers its binding energy and causes it to expand.
The net effect of these various processes is difficult to 
estimate without full $N$-body simulations.

\begin{figure*}
\centering
\includegraphics[scale=0.80,angle=-90.]{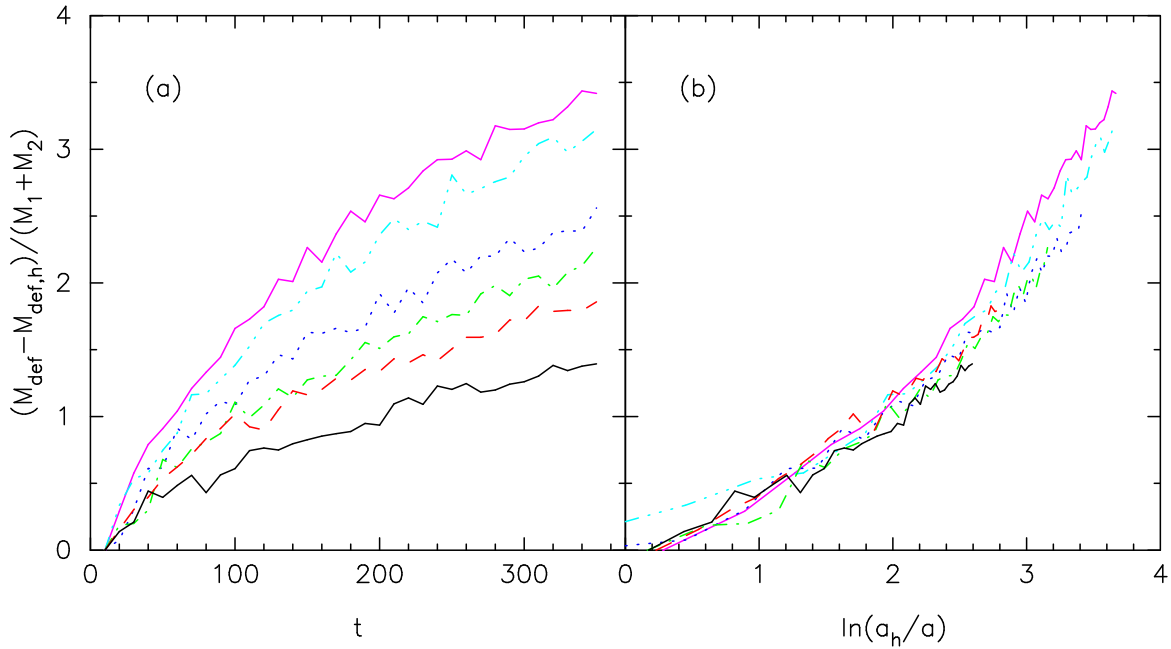}
\caption{Evolution of the mass deficits in the 
$N$-body integrations, vs. time (a) and semi-major  axis (b).
$a_h$ is the binary separation at $t=10$, when the
hard binary forms, and $M_{def,h}$ is the mass deficit
at  this time.
Line styles have the same meaning as in Figure~\ref{fig:av_ainv}.}
\label{fig:mdef}
\end{figure*}

\begin{figure}
\centering
\includegraphics[scale=0.4]{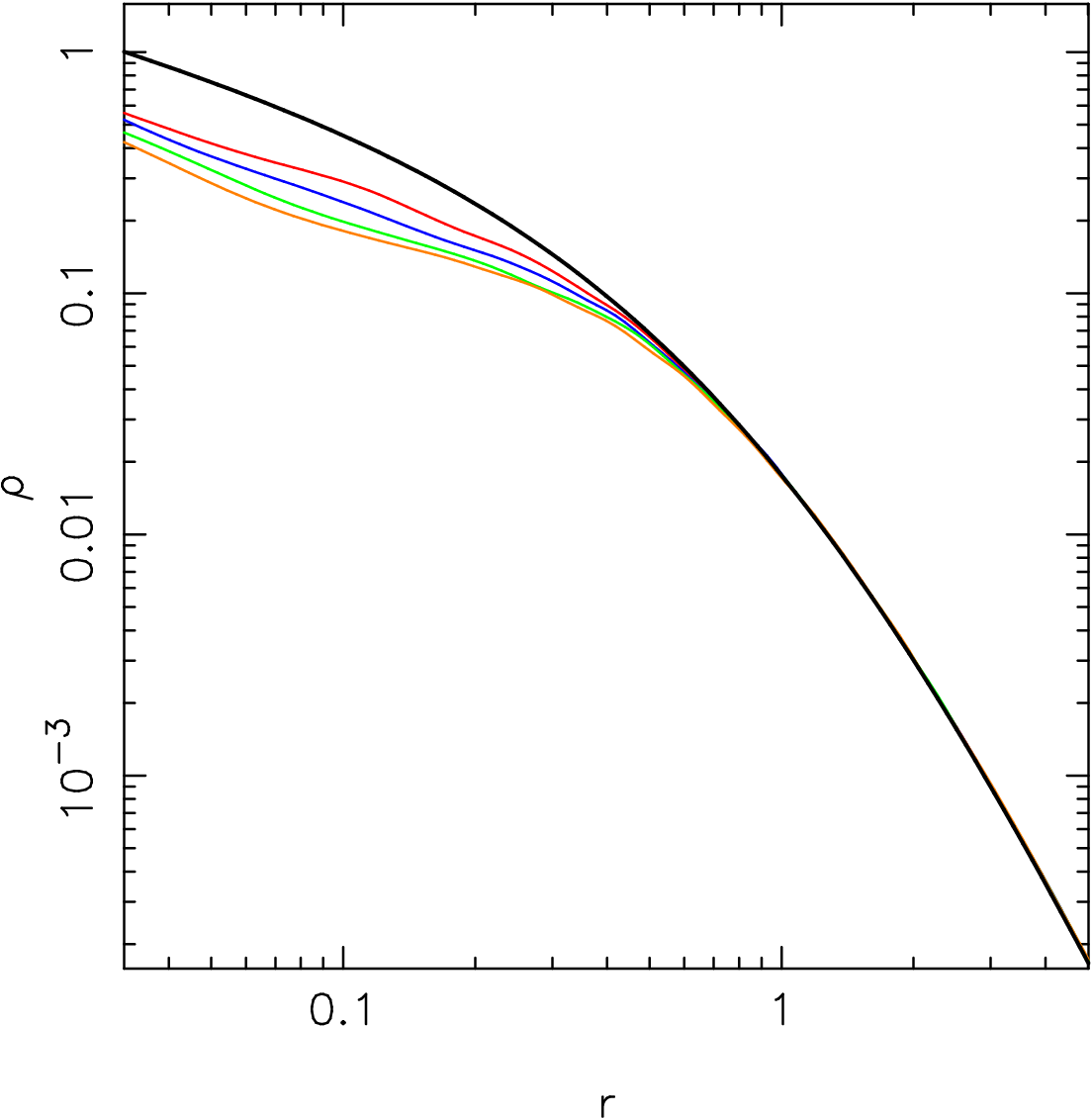}
\caption{Evolution of the mean density profile
in  the 64K integrations.
Black: $t=0$; red: $t=50$; blue: $t=150$; green: $t=250$;
orange: $t=350$.}
\label{fig:rhooft}
\end{figure}

We follow the standard practice of describing changes
in core mass in terms of the mass deficit $M_{def}$,
defined by \cite{MM:02} as the difference in integrated
mass between the density profile and the initial
density profile, within the region influenced by the binary.
Mass deficits have been estimated in a number galaxies
\citep{MM:02,Ravin:02,Graham:04,Merritt:06}
using assumed forms for the pre-existing density profile.
Figure~\ref{fig:mdef} shows $M_{def}$ versus time, and
versus binary semi-major axis, for the averaged 
$N$-body integrations.

As in previous work \citep{MM:01,Merritt:06}, we find that
the mass deficit increases suddenly when $a\approx a_h$,
to a value $M_{def}\approx\mtwo$.
Since the initial conditions adopted here are rather artificial --
neither of the black hole particles was placed at the center,
for instance -- the value which we find for $M_{def}$ at this time 
may not accurately reflect the value following a real
galaxy merger.
We therefore present in Figure~\ref{fig:mdef} $M_{def}-M_{def,h}$, 
the {\it change} in the mass deficit since the time at which
$a=a_h$;
as above, we take this time to be $t=10$
(Fig.~\ref{fig:av_ainv}).

When plotted vs. $a_h/a$ (Fig.~\ref{fig:mdef}b),
the $N$-dependence of the evolution almost disappears,
allowing the differential mass deficit to be expressed almost uniquely
in terms of the change in semi-major axis.
As shown below, a binary would not be expected to evolve past
$a^{-1}\approx 100 a_h^{-1}$ before gravitational wave losses
begin to dominate the evolution,
implying 
a maximum mass deficit of $\sim 5\mtwo$; however an extrapolation
of this prediction to the much larger $N$ regime of real
galaxies would be dangerous.
Figure~\ref{fig:rhooft} shows averaged density profiles
at various times for  the integrations with $N=65K$.

\section{The Fokker-Planck Model}

As shown in Figure~\ref{fig:fpslopes}, the $N$-dependence of
binary hardening rate in the $N$-body simulations
is $s\sim N^{-0.4}$.
This is substantially flatter than the $\sim N^{-1}$
dependence expected in a diffusively-repopulated
(``empty'') loss cone \citep{MM:03}, which makes it difficult
to extrapolate the $N$-body results to the regime of
real galaxies.
In this section we develop a Fokker-Planck model that
can reproduce the $N$-body results and which can also
be applied to systems with arbitrarily large $N$.
Unlike previous treatments of this problem based on encounter theory,
we allow the radial distribution
of matter to evolve in our Fokker-Planck models, 
due both to loss of stars that interact with the binary, 
and to diffusion in energy of non-interacting stars. 
These improvements will be shown to be crucial for accurately
reproducing the $N$-body results.
They also allow us, for the first time,
to make quantitative predictions about the 
evolution of the mass deficit in galaxies where binary evolution
is driven by collisional loss-cone repopulation.

\subsection{Loss-cone Dynamics}

Consider a spherical galaxy containing a massive central 
binary that acts like a sink, ejecting stars
that come sufficiently close to it.
Let $E=-v^2/2+\psi(r)$ be the binding energy per unit mass
of a star in the combined potential $\Phi(r)=-\Psi(r)$ of the galaxy
and the binary; the latter is approximated as $-G\mtwo/r$.
The binary defines a loss cone of orbits that satisfy
$J\lap J_{lc}(E)$, where
\beq
J^2_{lc}(E) = 2r_{lc}^2 \left[\psi(r_{lc})-E\right]\approx 2G\mtwo r_{lc};
\eeq
here $J$ is the angular momentum per unit mass of a star and 
$r_{lc}$ is the radius of the ejection sphere around the binary.

Suppose that the binary has interacted with and ejected
all stars that were initially on orbits satisfying $J\le J_{lc}$.
(In Fig.~\ref{fig:av_ainv}, this appears to have occurred by a time of
$\sim 15$.)
The binary's subsequent hardening is limited by the rate at which
stars are scattered onto previously depleted loss-cone orbits.
A fundamental quantity is the ratio 
$q_{lc}(E)$ between the orbital period $P(E)$ and the (orbit-averaged)
time scale for diffusional refilling of the consumption
zone (Paper I):
\beq
q_{lc}(E)\equiv \frac{1}{R_{lc}(E)} \oint 
\frac{dr}{v_r}
\lim_{R\rightarrow 0}
\frac{\langle(\Delta R)^2\rangle}{2R} .
\label{eq:qofe}
\eeq
Here $R\equiv J^2/J_c(E)^2$ is a dimensionless
angular momentum variable, $0\le R\le 1$, with
$J_c(E)$ the angular momentum of a circular orbit
of energy $E$, and $\langle\left(\Delta R\right)^2\rangle$ is
the diffusion coefficient associated with $R$.
The limit $R\rightarrow 0$ in equation~(\ref{eq:qofe}) 
reflects the approximation that only very eccentric
orbits are scattered into the binary; the
orbital period is likewise defined in terms of 
a $J=0$ orbit.
This approximation breaks down for the most bound orbits
but as we show, almost all of the loss cone repopulation
comes from stars weakly bound to the binary.

\begin{figure*}
\centering
\includegraphics[scale=0.7,angle=-90.]{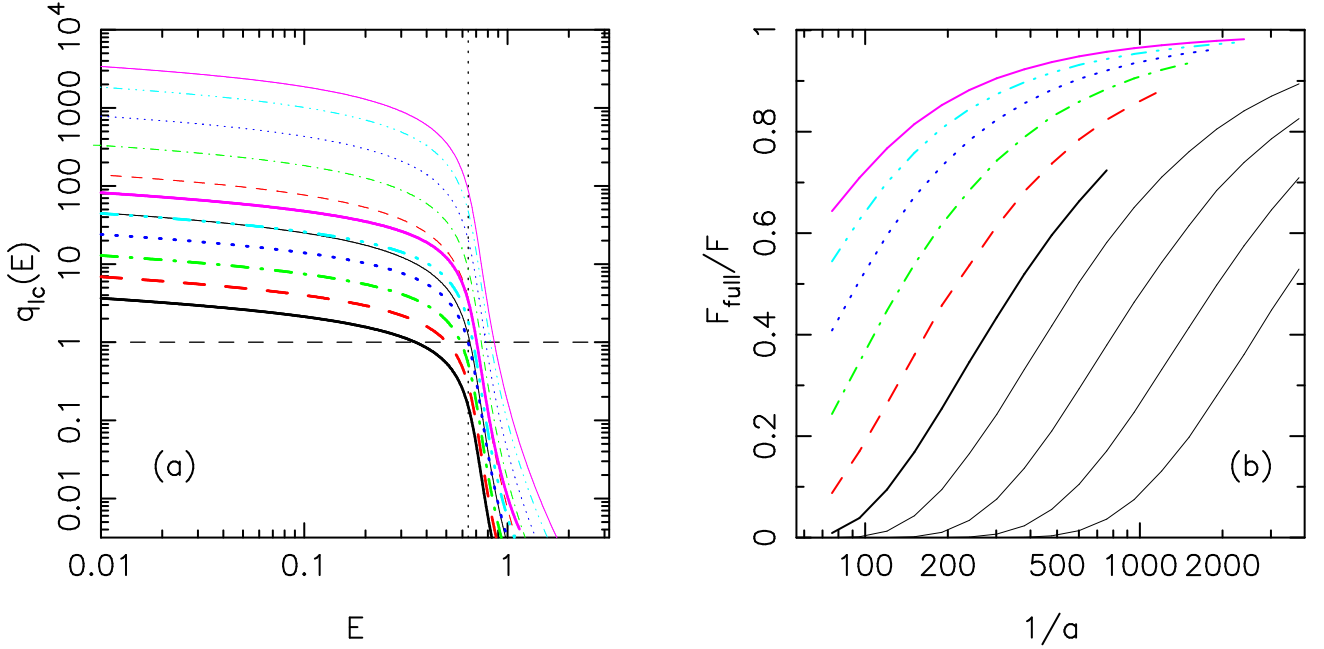}
\caption{(a) The function $q_{lc}(E)$ that describes the
ratio of the orbital period at $E$ to the timescale for 
diffusional refilling of the loss cone;
$q\gg 1$ indicates that the
loss cone is ``full,''  and real galaxies have $q_{lc}<1$.
Line styles have the same meaning as in Figure~\ref{fig:av_ainv}.
Thick curves show $q_{lc}$ for $a^{-1}=100$, when
binary has just entered the $N$-dependent phase of
its evolution (Fig.~6). 
Thin curves show $q_{lc}$ for the binary at the final time
step, $t=350$; 
the binary separation at this time is different for 
each $N$.
The radius of the loss sphere has been set to $a$.
Vertical dotted line is $E=\psi(r_h)$.
(b) The fraction of the flux of stars into the
binary's loss cone that is contributed at energies
where $q_{lc}\ge 1$, i.e., where the loss cone is
essentially full.
Lines show predictions for $N=(0.5,1,2,4)\times 10^6$.
These plots ignore binary-induced changes in the mass distribution
of the galaxy.
\label{fig:qofe}}
\end{figure*}

In the case of orbits with periods much shorter than the 
refilling time ($q_{lc}\ll 1$), the system is 
``diffusive'' and the loss cone is largely empty.
For orbits with periods much longer than the refilling
time ($q_{lc}\gg 1$), the system is in the ``pinhole'' or
``full loss cone'' regime.
In a galaxy containing a binary with fixed $r_{lc}$,
$q_{lc}$ increases with decreasing $E$, i.e. with
increasing distance from the binary.
The energy at which $q_{lc}=1$ is defined as the critical energy,
$E_{crit}$,
that separates empty- from full loss cone regimes.
The $N$-dependence of the problem appears via the
angular momentum diffusion coefficient 
$\langle\left(\Delta R^2\right)\rangle$,
which scales (approximately) linearly with the mean stellar mass,
i.e. inversely with $N$ for a fixed mass of the galaxy
(Paper I).
Other factors that influence $q_{lc}$ are the degree of central
concentration of the galaxy (high central density implies
larger $q_{lc}$) and the size $r_{lc}$ of the interaction sphere
(i.e. the binary semi-major axis).
Milosavljevic \& Merritt (2003) show that massive binaries
in real galaxies ($N\gap 10^9$) are essentially always in the 
empty loss cone regime, even in the extreme case of a 
$\rho\sim r^{-2}$ stellar density cusp, due to the
long relaxation times and to the large physical size
of a binary.

As a first step toward understanding the evolution of the
binary in our $N$-body simulations, we plot
in Figure~\ref{fig:qofe}a
$q_{lc}(E)$ for our initial galaxy model, assuming two values
for $r_{lc}$ at each $N$: $r^{-1}_{lc} =100$, corresponding
to the time $t\approx 15$ when the hardening rate has
just begun to exhibit a dependence on $N$
(Fig.~\ref{fig:av_ainv}); and $r_{lc} = a(t=350)$, 
the final value of $a$ (different for each $N$).
This  figure suggests that none of the integrations was fully
in the empty loss cone regime characteristic of real galaxies; 
even for $N=262k$, $q_{lc}>1$ except at energies close to $\psi(r_h)$
(as defined above, $r_h$ is the gravitational influence radius of the
central mass, i.e. the radius containing a mass in stars
equal to twice $\mtwo$).
As the binary hardens, $q_{lc}$ increases in all of the simulations,
and at the final time step, $q_{lc}>1$ at $E<\psi(r_h)$ for all
$N$, i.e. the binary has evolved essentially completely into
the full loss cone regime.

A more useful characterization of the binary's loss cone is
shown in Figure~\ref{fig:qofe}b.
For this figure, the flux of stars into $r_{lc}$ was
computed, 
and broken into two parts: the flux $F_{\rm full}$
originating from stars at energies such that $q_{lc}\ge 1$; and
$F_{\rm empty}$, from stars with energies such that 
$q_{lc}<1$.
The energy-dependent flux ${\cal F}(E)$ can be derived from the orbit-averaged
equation describing diffusion in $J$ (Eq.~19, Paper I):
\beq
{\partial {\cal N}\over\partial t}={R_{lc}\over P}q_{lc} {\partial\over\partial R}
\left(R{\partial {\cal N}\over \partial R}\right),
\label{eq:dNdt}
\eeq
where ${\cal N}(E,R,t)=4\pi^2P(E)J_c^2(E)f(E,R,t)$ is the
number density of stars in the $(E,R)$ plane.
\footnote{
We assume in writing equation~(\ref{eq:dNdt}) that the orbit-averaged
Fokker-Planck equation can be applied near the loss-cone boundary.
This is valid for the diffusively-repopulated loss cone of a 
binary in a real galaxy, but may not be valid at low energies 
in the $N$-body  simulations since the loss cone is nearly full and the
separation of time scales on which the orbit-averaging is
based breaks down.
Nevertheless equation~(\ref{eq:dNdt}) is traditionally
applied even in this regime \citep{CK:78,MT:99}.
Our expression for the flux does tend to the correct limit
in the full loss cone regime, $q_{lc}\gg 1$.
See Shapiro \& Marchant (1982) for a treatment of the
loss cone that is not based on the orbit-averaged approximation}.
The flux into the binary is
\begin{subequations}
\begin{eqnarray}
{\cal F}(E)dE &=& \left[-{d\over dt} \int_{R_0}^1 {\cal N}(E,R,t) dR\right]dE \\
&=& -{R_{lc}\over P} q_{lc} \left[R{d{\cal N}\over dR}\right]_{R_0}^1 dE \\
&=& 4\pi^2 J^2_{lc}(E) q_{lc}(E) \left[ R{\partial f\over\partial R}\right]_{R_0} dE.
\label{eq:FofE1}
\end{eqnarray}
\end{subequations}
In these expressions, $f$ has been allowed to fall to
zero at an angular momentum $R_0(E)$ that is different
from $R_{lc}(E)$.
Cohn \& Kulsrud (1979) derived an approximate expression for $R_0$:
\begin{equation*}
R_0(E) = R_{lc}(E) \times 
\begin{cases} 
\exp(-q_{lc}), & \text{$q_{lc}(E) > 1$}
\\
\exp(-0.186 q_{lc} -0.824 \sqrt{q}), & \text{$q_{lc}(E) < 1$.}
\end{cases}
\end{equation*}
For $q_{lc}\ll 1$, $R_0\approx R_{lc}$ but as $q_{lc}$ increases,
the loss cone is largely full and $R_0\approx 0$.
Finally, we adopt the steady-state solution to
equation~(\ref{eq:dNdt}) for $f$, i.e.
\beq
f(R;E) = {\ln\left(R/R_0\right)\over\ln\left(1/R_0\right)-1} 
\overline{f}(E)
\label{eq:fofRE}
\eeq
(assuming $R_0\ll 1$) implying a diffusive flux
\beq
{\cal F}(E)dE = 4\pi^2 J^2_{lc}(E) q_{lc}(E) {\overline {f}(E)\over\ln\left(1/R_0\right)-1} dE.
\label{eq:FofE2}
\eeq
Here, $\overline{f}=\int_0^1 f(E,R) dR$ is the isotropic
$f$ that has the same total number of stars at each $E$
as the true $f(E,R)$.

As noted above, the loss cone of a binary black hole
in a real galaxy is essentially empty, i.e. almost all
of the stars scattered into the binary would come
from energies $E>E_{crit}$.
Figure~\ref{fig:qofe}b shows the results of applying
equation~(\ref{eq:FofE2}) to our initial $N$-body model,
 with $r_{lc}=a$ and with
$a$ allowed to vary over the range $100\le a^{-1}\le a^{-1}_{t=350}$.
In this figure,
$F_{full}$ is the flux integrated from $0$ to $E_{crit}$
and $F$ is the total flux.
At the start of the $N$-body integrations,
Figure~\ref{fig:qofe}b suggests that the binary
in the larger-$N$ models ($N\gap 64k$) is essentially
in the empty loss cone regime, $F_{full}\ll F$.
However by the final time step, the binary has shrunk
and entered into the ``pinhole'' regime, $F_{full}>F_{empty}$,
for all $N$.
In the integrations with $N\lap 16k$,
the binary is in the full loss cone regime 
from the start.

Figure~\ref{fig:qofe}b also includes curves for the cases
$N=(0.5,1,2,4)\times 10^6$.
Values of $N$ up to $4\times 10^6$ are now computationally
feasible via direct-summation codes combined with 
special-purpose hardware \citep{Harfst:06}, and Figure~\ref{fig:qofe}b
suggests that this $N$ value is large enough to place the
binary effectively in the empty loss cone regime for
most of its evolution.
(The minimum required $N$ would be larger than this if the binary
were given the smaller mass, $\sim 10^{-3}M_{gal}$, typical
of black holes in real galaxies, or if the galaxy model
were more centrally concentrated.)

Figure~\ref{fig:qofe} illustrates the difficulty of 
scaling the binary evolution observed in our $N$-body simulations
to real galaxies.
In the empty loss cone (i.e. diffusive, large-$N$) limit,
the supply of stars to the binary scales as 
$\langle\left(\Delta R^2\right)\rangle\propto m_\star\propto N^{-1}$
for a fixed total galaxy mass (ignoring the weak dependence of the
Coulomb logarithm on $N$).
In the full loss cone (pinhole, small-$N$) limit,
the loss cone flux is independent of $N$.
In between these limiting cases, one expects (Paper I)
that the flux, and hence the hardening rate of the binary,
scales as $\sim N^{-\beta},\ 0<\beta<1$.
This is consistent with the $s\sim N^{-0.36}$ 
dependence observed here (equation~\ref{eq:sofN}).
Figure~\ref{fig:qofe}b suggests that of order $N\approx 10^7$ 
stars would be required before the binary is
comfortably in the empty loss cone regime, allowing
its evolution to be reliably scaled to larger values of
$N$.

Even if we were in this regime, the expressions given above
for  the flux of stars into the binary's loss cone
might not accurately predict the binary's
evolution, since they ignore changes in the galaxy's structure.
Figure~\ref{fig:rhooft} suggest that these changes
are significant: the density near the galaxy's center
changes by a factor $\sim 2$ as the binary hardens.
We now consider a model that includes
both changes in the binary due to interaction with stars,
as well as binary-induced changes in the stellar distribution,
and that can be reliably scaled to the large-$N$ regime
of real galaxies.

\subsection{Evolutionary Model}

The $J$-directed flux of stars into the binary,
described by equation~(\ref{eq:FofE2}),
implies a decrease in the number of stars 
at $J_{lc}\lap J \lap J_c(E)$.
In Paper I, this decrease was followed 
by integrating equation~(\ref{eq:dNdt}) forward in time
at fixed $E$.
The justification for treating the problem in this 
restricted way was the
difference in time scales between $E$- and $J$-diffusion;
the former occurs in a time $\sim\tr$ while the latter 
requires $\sim (a_h/r)\tr$.
The evolution of $f(J;E)$ and ${\cal F}(E)$ 
over the shorter of these time scales
was followed starting from a completely emptied loss cone
and the change in the density of the core was computed from
the changes in $N(J;E)$ at every $E$.

In the present paper, we focus on changes that take
place over the longer of these two time scales, $\sim\tr$.
This allows us to largely ignore the initial conditions,
and to assume that an expression like~(\ref{eq:fofRE})
is an adequate description of the $J$-dependence of $f$
at every $E$.
However it also implies that we can not ignore changes
in $E$, which occur on timescales of $\sim\tr$.

\begin{figure*}
\centering
\includegraphics[scale=0.8,angle=-90.]{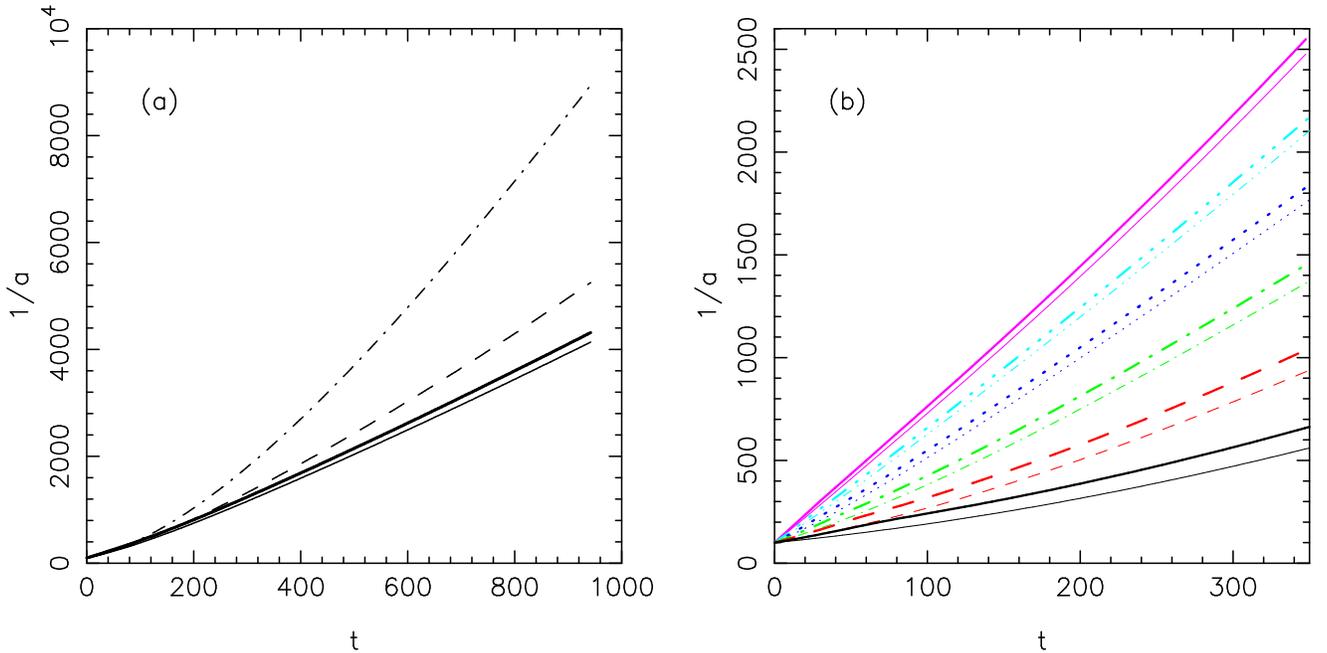}
\caption{(a) Fokker-Planck evolution of binary semi-major axis in
a set of integrations designed to mimic the $N$-body
simulations with $N=65$k.
{\it Dot-dashed line:} fixed potential and density;
{\it dashed line:} fixed potential, evolving density;
{\it thin solid line:} evolving density and potential, no re-ejections;
{\it thick solid line:} evolving density and potential with re-ejections.
(b) Fokker-Planck integrations with parameters chosen to mimic
the $N$-body simulations with various $N$; color coding is the same
as in the $N$-body figures above.
{\it Thin lines:} no re-ejections;
{\it solid lines:} with re-ejections.
}
\label{fig:fp}
\end{figure*}

On these longer time scales, the evolution of the
density near the binary is a competition between loss of stars
that diffuse onto low-$J$ orbits and are ejected by the binary,
as described by ${\cal F}(E)$, 
and replenishment due to  stars that
diffuse in energy from regions of lower $E$, i.e. larger radius.
Beyond a certain radius, the relaxation time is so
long that the $E$-directed flux can not compensate for the integrated
loss-cone flux, $\int {\cal F}(E) dE$, and
the mean density within this radius must drop -- implying
the creation of a mass deficit.

We can approximate the evolution of the galaxy/binary system
in this late-time regime via a modification of the orbit-averaged 
Fokker-Planck equation for $f(E)$:
\beq
{\partial N\over\partial t} = -{\partial F_E\over\partial E}
- {\cal F}(E,t),
\label{eq:dNdt2}
\eeq
where ${\cal F}$ is the $J$-directed flux defined in
equation~(\ref{eq:FofE2}),
and $F_E$ is the energy-directed flux, given by
\begin{subequations}
\begin{eqnarray}
F_E &=& -D_{EE}{\partial f\over\partial E} - D_Ef \\
D_{EE} &=& 64\pi^4G^2m_\star\ln\Lambda
\bigg[q(E) \int_0^E dE'f(E') \nonumber \\
&+&  \int_E^\infty dE' q(E')f(E')\bigg], \\
D_E &=& -64\pi^4G^2m_\star\ln\Lambda\int_E^\infty dE'p(E')f(E')
\end{eqnarray}
\label{eq:FP2}
\end{subequations}
\noindent 
In these expressions, $f(E)$ is understood to be the
{\it mass} density of stars in phase space associated with the function 
$\overline{f}(E)$ defined above, and the quantities
$F_E$ and ${\cal F}$ are mass fluxes.
$N(E)dE=4\pi^2p(E)f(E)dE$ is energy-space distribution,
with $p(E)$ and $q(E)$ the phase-space weighting factors,
\begin{subequations}
\begin{eqnarray}
p(E) &=& 4\int_0^{r_{max}(E)} v(r)r^2 dr, \\
q(E) &=& {4\over 3} \int_0^{r_{max}(E)} v^3(r)r^2 dr,
\end{eqnarray}
\end{subequations}
and $v=\left[2\Phi(r)-2E\right]^{1/2}$.
Near the binary, where the potential is close to Keplerian,
$p(E)\approx 2^{-3/2}\pi G^3\mtwo^3|E|^{-5/2}$
and $q(E)=(2^{1/2}\pi/6)G^3\mtwo^3|E|^{-3/2}$.
$\ln\Lambda\approx \ln(\mtwo/m_\star)$ is the Coulomb logarithm.
Henceforth $f$ and $N$ are explicitly defined 
as mass (not number) densities,
and ${\cal F}$ is the mass flux into the binary's loss cone.

An equation like (\ref{eq:dNdt2}), in which
the $J$-dependence of $f$ is contained 
implicitly in ${\cal F}(E,t)$, 
was first written by Bahcall \& Wolf (1977).
It has since been adopted by a number of other authors
to describe the evolution of the distribution
of stars, compact objects or dark matter around a single
supermassive black hole \citep{MCD:91,Merritt:04,Hopman:06}.
It is being used for the first time in the present paper
to describe the evolution of the stellar distribution
about a binary black hole.
Since ${\cal F}$ scales only as $\sim\log r_{lc}^{-1}$ 
(equation~\ref{eq:FofE2}),
the ratio of the two terms on the  right hand side of 
equation~(\ref{eq:dNdt2}) is not greatly affected by the
much greater (in linear extent) size of the loss cone
of a binary compared with a single black hole.

The relation between the flux into the binary's loss cone
and the rate of change of its semi-major axis $a$ is
\beq
{d\over dt} \left({GM\mu\over 2a}\right) = -\int {\cal F}(E,t)\Delta E dE 
\eeq
with $\mu\equiv M_1M_2/M$, the binary reduced mass and 
$\Delta E (E)$ the mean specific energy change of stars,
originally at energy $E$, that interact with the binary.
In Paper II, we set
\begin{subequations}
\begin{eqnarray}
\Delta E &=& \Delta E_{\rm Hills} = -\langle C\rangle {G\mu\over a}, 
\label{eq:harda}\\
s(t) &\equiv& {d\over dt}\left({1\over a}\right) = {2\langle C\rangle\over aM}
\int {\cal F}(E,t) dE.
\label{eq:hardb}
\end{eqnarray}
\label{eq:de1}
\end{subequations}
The coefficient $\langle C\rangle$ is independent of energy
for stars that interact with a ``hard'' binary
\citep{Hills:83,Mikkola:92,Quinlan:96a}.
Hardness is defined as $V_{bin}/\sigma$, where
$V_{bin}=\sqrt{GM_{12}/a}$ is the relative velocity of the components
of the binary
and $\sigma$ is the stellar velocity dispersion in the unperturbed
galaxy.
An equal-mass binary is in the ``hard'' regime when $V_{bin}/\sigma\gap 3$
\citep{Quinlan:96a}.
In the current models, 
\beq
{V_{bin}\over\sigma_p} \approx 0.36 \left({1\over a}\right)^{1/2}
\eeq
with $\sigma_p\approx 0.278$ the peak velocity dispersion 
in the $\gamma=0.5$ Dehnen model.
The $N$-dependent phase of binary evolution begins at
$a^{-1}\approx 100$ in the $N$-body models
(Fig.~\ref{fig:av_ainv}), hence
$V_{bin}/\sigma_p\gap 3.6$ and equation~(\ref{eq:de1})
is expected to be accurate.
In Paper II, setting $\langle C\rangle\approx 1.25$ 
was found to reproduce the $N$-body hardening rates.
\cite{Yu:02} argued for a similar value of $\langle C\rangle$.

\begin{figure*}
\centering
\includegraphics[scale=0.8,angle=-90.]{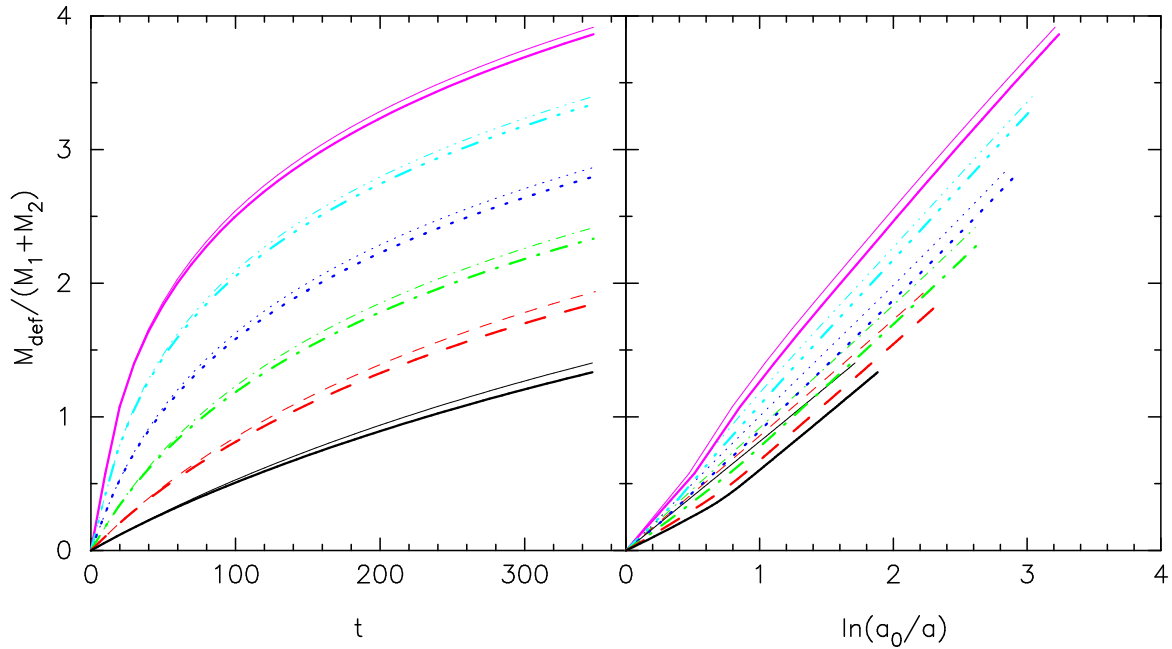}
\caption{Evolution  of the mass deficit in the suite of
Fokker-Planck integrations presented in Fig.~\ref{fig:fp}.
Line styles have the same meaning as in Figure~\ref{fig:av_ainv}.
}
\label{fig:fpmdef}
\end{figure*}

Expressions like (\ref{eq:harda}) were derived from scattering
experiments that allowed for the possibility of multiple
interactions between star and binary.
However the confining effect of the galaxy's gravitational
potential was ignored.
As noted in \cite{MM:03},
stars ejected once by the binary can interact with it
again as they return to the nucleus on nearly-radial orbits.
If the energy change during the first interaction is not large enough 
to eject the star completely from the galaxy, it will experience 
one or more ``secondary slingshots'', and the total
energy extracted from the binary by the star will be the sum of the discrete
energy changes during the interactions.

A minimum condition for re-ejection is that a star remain
bound to the galaxy after its first interaction with the binary,
$E+\Delta E\gap 0$.
Most stars that interact with the binary have apocenters $\sim r_h$
\citep{MM:03}; since the gravitational potential at this radius
is dominated by the galaxy, we can write this condition for re-ejection
as $|\Delta E| \lap \Phi_\star(0)$
with $\Phi_\star(0)$ the central value of the galaxy's (stellar)
gravitational  potential.
The $\gamma=0.5$ Dehnen models used here have $\Phi_\star(0)\approx 0.67$
in the adopted units, implying $a^{-1}\lap 200$ for re-ejection.

Even if a star satisfies this condition, 
re-ejection will only be effective if the star
remains in the binary's loss cone for longer than an orbital period, i.e.
if $q(E)\lap 1$.
Re-ejection will also fail for a star with apocenter greater
than some $r_{max}\gg r_h$, since the overall potential in a real
galaxy is never precisely
spherical and the star will be perturbed from its nearly radial
orbit on the way in or out \cite{Vicari:07}.

We considered a modified form of equation~(\ref{eq:de1}) that
accounts for re-ejections.
Let $\Delta E_{\rm max} = \Phi(r_{max})-\Phi(r_h)$.
Re-ejection was assumed to occur if the following conditions
were both satisfied:
(i) $\left|\Delta E_{\rm Hills}\right| < 
     \left|\Delta E_{max}\right|$; 
(ii) $q(E+\Delta E_{\rm Hills}) < q_{max} \approx 1$.
Condition (i) guarantees that the star remains bound
to the galaxy after the first ejection, with apocenter
$\le r_{max}$.
This condition is roughly equivalent to
$G\mu/a<\Phi_{gal}(0)\approx 1$ and is satisfied for
$a^{-1}\lap 500 \approx 5 a_h^{-1}$ in our models, 
i.e. during the early phases of binary evolution.
Condition (ii) guarantees that the star
will remain within the binary loss cone for of order
one orbital period or longer after the first ejection; 
this condition is satisfied at large (i.e. bound) values of $E$
(Fig. 13).

For $a^{-1}$ greater than $\sim 160$, 
which occurs shortly after formation of a hard binary (Fig.~5), 
even a single re-ejection would give a star enough
energy to escape the galaxy.
Hence, at energies such that conditions (i) and (ii) 
were both satisfied,
we set $\Delta E=2\Delta E_{\rm Hills}$,
while if either condition was not satisfied,
re-ejection was assumed not to occur and we set 
$\Delta E = \Delta E_{\rm Hills}$.
This scheme has two parameters, $q_{max}$ and $r_{max}$;
the results are weakly dependent on $r_{max}$ for
$r_{max}\gg r_h$ and we fixed $r_{max}=100 r_h$.
The consequences of varying $q_{max}$ are discussed below.

Finally, we need to account for changes in the gravitational
potential as the stellar distribution evolves.
(We ignore possible changes in the mass of the binary.)
Here we follow H\'enon's (1961) scheme of assuming that
$f$ remains a fixed function of the radial adiabatic
invariant as the potential is adjusted.
Our numerical schemes for advancing $f$ was based closely
on the algorithms described by \cite{Cohn:80} and 
\cite{Quinlan:96b}.

\subsection{Comparison with the $N$-Body Integrations}

Figure~\ref{fig:fp}(a) compares the evolution of $a^{-1}$
in a set of Fokker-Planck integrations with initial
conditions chosen to mimic those in the $N$-body integrations
($\gamma=0.5$, $M=0.01$, $a^{-1}(t=0)=0.01$).
Fixing $\rho(r)$ and $\Phi(r)$ (dot-dashed curve)
is equivalent to the assumptions made by Yu (2002),
who ignored changes in the stellar distribution as the
binary evolved.
Allowing the density and potential to evolve (solid lines) 
results in a considerably lower hardening rate for the binary.
Including the secondary-slingshot (heavy solid line) increases
the hardening rate but only slightly; as explained above,
once the binary becomes hard, most stars that interact with
it are ejected completely from the galaxy and do not
return to the binary's sphere of influence.

Figure~\ref{fig:fp}(b), which can be compared with
Figure~\ref{fig:av_ainv}(b), shows the evolution of binary 
semi-major axis in a set of Fokker-Planck integrations
with the same values of $N$ as in the $N$-body integrations.
The correspondence is quite good; the Fokker-Planck
integrations show a slightly steeper dependence of the
binary hardening rate on $N$ (Fig.~\ref{fig:fpslopes}).
The evolution of the mass deficit as derived from the Fokker-Planck
integrations is shown in Figure~\ref{fig:fpmdef}
(cf. Fig.~\ref{fig:mdef}).
Here the correspondence is not quite as good, but
still reasonable; the weak dependence of $M_{\rm def}$
on $N$ for large $N$ is well reproduced.

\section{Predictions of the Fokker-Planck Model for Large $N$}

Having established that the Fokker-Planck model 
can mimic the joint binary/galaxy evolution seen
in the $N$-body integrations, for various values of
$N\lap 10^5$, we now extend this model to the 
much larger $N$ regime of real galaxies.
The goal is both to predict the long-term evolution of a massive
binary in a real galaxy, and also to record the 
changes in the central structure of the galaxy.

\begin{figure*}
\centering
\includegraphics[scale=0.75,angle=-90.]{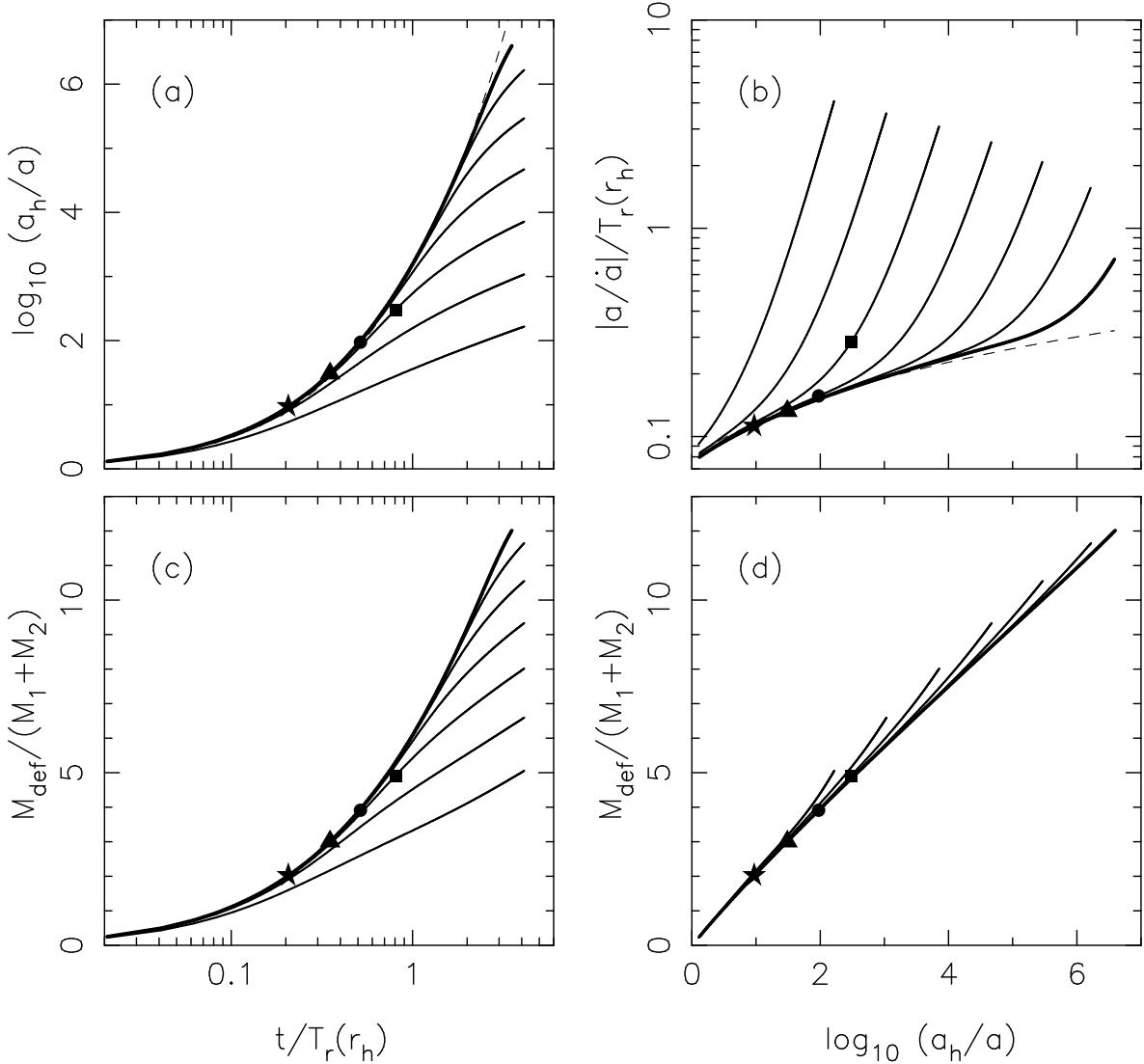}
\caption{Joint binary-galaxy evolution in Fokker-Planck models
with $M_1=M_2$ and $M_{12}=10^{-3}M_{\rm gal}$.
(a) Binary semi-major axis; (b) binary hardening time;
(c) mass deficit as a function of time, and (d) mass deficit
as a function of binary separation.
Different lines correspond to different values of 
$N\equiv M_{\rm gal}/m_\star$: $N=10^6,10^7,...,10^{11}, 10^{12}$
(thick line).
Symbols mark the time $t_{eq}$ at which the binary hardening rate equals
the gravitational radiation evolution rate,
assuming a binary mass of $10^5\msun$ (squares), 
$10^6\msun$ (circles), $10^7\msun$ (triangles), and $10^8\msun$ (stars).
Filled symbols denote models in which $N$ is roughly equal to its
value in real galaxies, for each value of $\mh$.
Dashed lines in panels (a) and (b) are the analytic model described
in the text.
}
\label{fig:0.5_1.0}
\end{figure*}

Results for a galaxy containing a binary with
$\mtwo\equiv M_1+M_2=10^{-3}M_{\rm gal}$ and two mass ratios 
$\q\equiv M_2/M_1=(1,0.1)$
are shown in Figures~\ref{fig:0.5_1.0} and \ref{fig:0.5_0.1} 
respectively.
The initial galaxy model was a Dehnen sphere,
equation~(\ref{eq:dehnen}), with $\gamma=0.5$.
This is the shallowest central slope that is consistent
with an isotropic phase-space distribution around a central
point mass; it is also a fair representation of the
core profiles that are produced during the ``rapid'' phase of cusp
destruction that accompanies the initial formation of the
massive binary \citep{MS:06}.
Fokker-Planck integrations were carried out for different
values of $N\equiv M_{\rm gal}/m_\star = (10^6, 10^7,...,10^{12})$.
The time axis in these plots is the relaxation time measured
at the binary's influence radius in the initial model;
all integrations were continued until $t=4\tr(r_h)$.
Equation~(\ref{eq:ah}) was used to set the initial value
of $a$; quantities like the mass deficit in Figures~\ref{fig:0.5_1.0}
and~\ref{fig:0.5_0.1} should be interpreted as the accumulated change
in these quantities after the binary first
becomes ``hard.''
Unless otherwise stated, re-ejections were ignored.

In all of the integrations, the binary begins in the
diffusive, or empty loss cone, regime ($q_{lc}\gg 1$) 
due to its large initial separation, and evolves toward the 
pinhole, or full loss cone, regime ($q_{lc}\gap 1$) as it hardens.
The transition to the pinhole regime occurs later for larger $N$; 
for $N=10^{12}$ (the heavy curves in Figs.~\ref{fig:0.5_1.0} 
and~\ref{fig:0.5_0.1}) the binary remains essentially in the diffusive
regime until the end of the integration at $4\tr(r_h)$.
However we argue below that evolution of binaries in real galaxies
would typically be expected to terminate before the pinhole regime is reached.

\subsection{Binary Hardening Rates}

\begin{figure*}
\centering
\includegraphics[scale=0.75,angle=-90.]{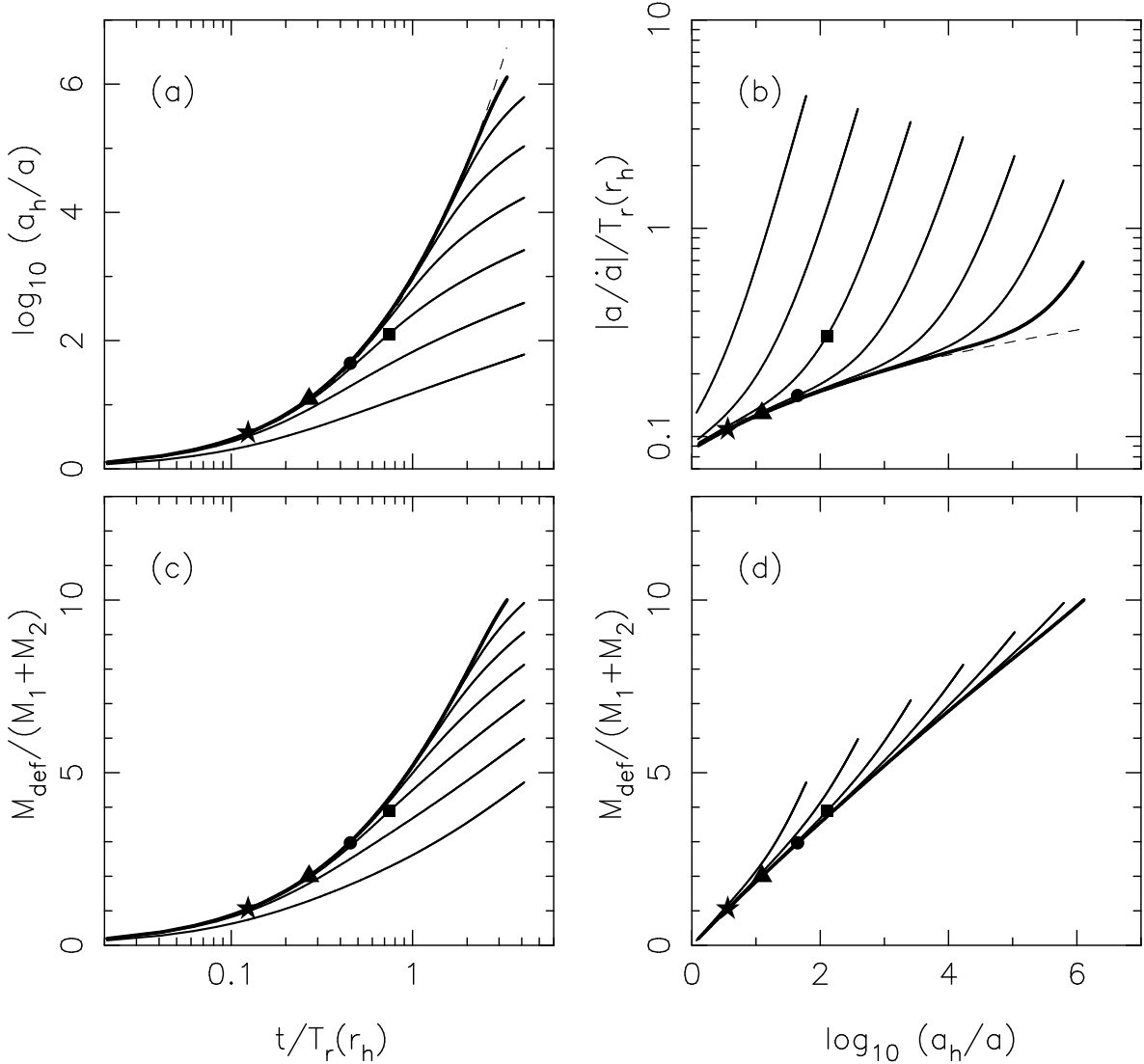}
\caption{Like Fig.~\ref{fig:0.5_1.0} but for
$M_2=0.1M_1$ and $M_{12}=10^{-3}M_{\rm gal}$.
}
\label{fig:0.5_0.1}
\end{figure*}

Figures~\ref{fig:0.5_1.0} and~\ref{fig:0.5_0.1} show that 
at large $N$, the binary hardening time,
\beq
\thard\equiv \left|{a\over\dot a}\right|,
\eeq
 tends to a fixed
fraction of $\tr(r_h)$ at any given $a$.
This is the ``empty loss cone'' regime.
The ratio $\thard/\tr(r_h)$ increases from $\sim 0.1$ at large $a$,
i.e. early times, 
to $\sim 0.3$ when $a\approx 10^{-5}a_h$, with a weak
dependence on binary mass ratio.
We will argue below that binary black holes in real
galaxies lie close to the large-$N$ hardening curves 
throughout much of their evolution and so it is of interest
to develop an analytic understanding of this regime.

Since $q_{lc}(E)$ (equation~\ref{eq:qofe})
is the ratio of the orbital period to the 
diffusional loss cone refilling time at energy $E$, i.e. 
$q_{lc}(E)\approx P(E)/[R_{lc}(E)\tr(E)]$,
we can rewrite the flux of stars into the binary, 
equation~(\ref{eq:FofE2}), as
\beq
{\cal F}(E)dE \approx 4\pi^2J_c^2(E)P(E)\tr^{-1}(E) 
{\overline {f}(E)\over\ln\left(1/R_0\right)-1} dE.
\label{eq:FofE3}
\eeq
Assuming a fixed mass model for the galaxy, 
the flux into the binary, integrated over one relaxation time, 
scales therefore as
\begin{subequations}
\begin{eqnarray}
{\cal F}(E)\tr(E) &\propto& \left[\ln\left(1/R_0\right)-1\right]^{-1} \\
&\approx& \left[\ln R_{lc}^{-1}\right]^{-1},\ \ q_{lc}\ll 1; \\
&\approx& q_{lc}^{-1},\ \ \ \ \ \ \ \ \ \ \ \ \ \ q_{lc}\gg 1.
\end{eqnarray}
\end{subequations}
The binary hardening rate is fixed by ${\cal F}$
and $a$ (equation~\ref{eq:hardb}), so these expressions imply that the 
binary's evolution over a specified  number of relaxation
times will be smaller for smaller $N$, i.e. larger $q_{lc}$;
while in the large-$N$ limit, the evolution rate at a given $a$
will be determined solely by $\tr$.
These predictions are consistent with the upper panels of
Figures~\ref{fig:0.5_1.0} and~\ref{fig:0.5_0.1}.

The gradual decrease with time of the hardening rate
is due to two factors: the decreasing size of the binary,
and the declining density of the core.
Again ignoring changes in the core structure, 
the expressions given above can be used
to estimate how the hardening rate varies with $a$.
The result, in the large-$N$ limit, is
\beq
{1\over \tr}\left|{a\over\dot a}\right| \equiv {\thard\over\tr}
\propto \ln\left({a_h\over a}\right),
\eeq
i.e. the fractional change in $a$ over one relaxation time
is weakly dependent on $a$ for large $N$.

We tried fitting a similar function to the large-$N$ hardening
curves in Figures~\ref{fig:0.5_1.0} and~\ref{fig:0.5_0.1}, i.e.
\beq
{1\over \tr}\left|{a\over\dot a}\right| = A\ln\left({a_h\over a}\right) + B.
\label{eq:anal}
\eeq
The results are shown as the dashed lines in 
Figures~\ref{fig:0.5_1.0}b and~\ref{fig:0.5_0.1}b.
We found good fits for
\beq
A=(0.016,0.017), \ \ \ \ \ \ B=(0.08,0.09)
\label{eq:aandb}
\eeq
for $\q=(1,0.1)$ respectively.
The weak dependence of the fitting parameters on
binary mass ratio reflects the lack of a
mass ratio dependence in the evolution equations
(\ref{eq:de1}).

Integrating equation~(\ref{eq:anal}) gives a simple expression
for the time dependence of the binary semi-major axis:
\beq
\ln\left({a_h\over a}\right) = -{B\over A} + \sqrt{{B^2\over A^2}
+ {2\over A}{t\over \tr(r_h)}}
\label{eq:tdepend}
\eeq
where $t$ is defined, as in Figures~\ref{fig:0.5_1.0} and
\ref{fig:0.5_0.1}, as the time since the binary first became
hard, i.e. the time since $a=a_h$.
This function is plotted in Figures~\ref{fig:0.5_1.0}a 
and~\ref{fig:0.5_0.1}a, where it again provides an excellent
fit to the large-$N$ evolution curves.

We now show that real black hole binaries are expected to
be in this empty loss cone regime throughout most or all of their evolution.
Maximum traversal of the tracks in Figures~\ref{fig:0.5_1.0}
and~\ref{fig:0.5_0.1} will occur if no physical process,
aside from interactions with stars, affects the hardening rate
until the gravitational radiation regime is reached.
The time scale associated with gravitational radiation 
is \citep{Peters:64}
\beq
\tgr = \left|{a\over\dot a}\right|_{\rm gr} = 
{5\over 64}{c^5\over G^3} {a^4\over \mu M_{12}^2}
\label{eq:peters}
\eeq
where $\mu\equiv M_1M_2/M_{12}$ is the reduced mass of the binary
and a circular orbit has been assumed.
Following \cite{Living}, 
$\tgr$ can be expressed in terms of $\mh\equiv M_{12}$
and $a_h$ using the $\mh-\sigma$ relation, equation~(\ref{eq:msigma}),
as
\beq
\tgr\approx 5.7\times 10^{10} {\rm yr} {\q^3\over (1+\q)^6} 
M_{\bullet,6}^{-0.65}\tilde{a}_{-2}^4
\label{eq:tgr}
\eeq
with $\tilde{a}\equiv a/a_h$ and
$\tilde{a}_{-2}=a/(0.01 a_h)$.

We define $t_{eq}$ as the time when $\thard= \tgr$.
In order to extract $t_{eq}$ in physical units from the Fokker-Planck
integrations, we need to assign a value in years to $\tr(r_h)$.
This we do via the straight-line fit to the data in
Figure~\ref{fig:trrh}.
Combining equations~(\ref{eq:trsigma}), (\ref{eq:anal}), 
and (\ref{eq:tgr}) 
the condition $\thard=\tgr$ becomes
\beq
\left({a_h\over a}\right)^4\left[A\ln\left({a_h\over a}\right)+B\right] = 7.1\times 10^8 \q^3\left(1+\q\right)^{-6} M_{\bullet,6}^{-2.19}
\label{eq:noname}
\eeq
with $M_{\bullet,6}\equiv \mh/10^6 M_\odot$.
We define $a_{\rm eq}$ as the value of $a$ that satisfies this equation.
For $\mh=(10^5,10^6,10^7,10^8)M_\odot$,
i.e. $\sigma\approx (44,70,112,180)$ km s$^{-1}$,
and using the values of $A$ and $B$ derived above,
equation~(\ref{eq:noname}) implies
\beq
a_h/a_{\rm eq} \approx (315,93,27,8.0)
\eeq
for $\q\equiv M_2/M_1=1$,
and 
\beq
a_h/a_{\rm eq} \approx (140,40,12,3.5)
\eeq
for $\q=0.1$.
The corresponding times are
\begin{eqnarray}
t_{eq} &\approx& (0.73,0.53,0.35,0.20) \times \tr(r_h) (\q=1), \\
       &\approx& (0.65,0.54,0.27,0.13) \times \tr(r_h) (\q=0.1).
\end{eqnarray}

The values just computed for $a_{eq}$ and $t_{eq}$ correspond
to the large-$N$ (empty loss cone) limit of the Fokker-Planck
equation, i.e. to the heavy curves in 
Figures~\ref{fig:0.5_1.0} and~\ref{fig:0.5_0.1}.
The filled symbols in those figures
show where $\thard=\tgr$ on the four tracks that
best correspond to the four values just considered for $\mh$.
Since $\mh\approx 1\times 10^{-3} M_{gal}$
\citep{MF:01},
we set $N= (10^8,10^9,10^{10},10^{11})$
for $\mh= (10^5,10^6,10^7,10^8)\msun$.
The symbols confirm that binary black holes of
mass $\mh\gap 10^{5.5} M_\odot$ remain essentially 
in the empty loss cone regime throughout their evolution.
For binaries of mass $\mh = 10^5 \msun$,
the evolution just prior to the gravitational radiation
regime begins to depart from that of a diffusive loss cone,
resulting in somewhat lower hardening rates than
predicted by equations~(\ref{eq:anal}) and (\ref{eq:tdepend}).
The discrepancy with the analytic expressions would be
expected to increase still more for binaries of still
lower mass (if such exist).

\subsection{Mass Deficits}

Next we consider the effect of the binary on the structure
of the galaxy's core.
Evolution of mass deficits is plotted in the lower panels of 
Figures~\ref{fig:0.5_1.0} and~\ref{fig:0.5_0.1}.
Particularly striking are the panels showing $M_{\rm def}$
vs. binary hardness, $a_h/a$.
As was true in the $N$-body integrations, long-term
evolution of the binary generates mass deficits that are 
very well predicted by the change in binding energy of the 
binary black hole, i.e. by $a_h/a$.
This dependence is accurately described by
\beq
{M_{\rm def}\over M_{12}} \approx (1.8, 1.6) \log_{10}\left(a_h/a\right)
\eeq
where the numbers in parentheses refer to 
$\q=(1,0.1)$ respectively.
The mass deficits generated between formation
of a hard binary, and the start
of the gravitational radiation regime, are given by
setting $a=\aeq$ in this expression, i.e.
\begin{eqnarray}
M_{\rm def} &\approx& (4.5,3.5,2.6,1.6)M_{12}\ \ (\q=1) \\
            &\approx& (3.4,2.6,1.7,0.9)M_{12}\ \ (\q=0.1).
\end{eqnarray}
for $M_{12}=(10^5,10^6,10^7,10^8)\msun$.
These values should be added to the mass deficits $M_{\rm def,h}$
generated during the rapid phase of binary formation,
i.e. $M_{\rm def,h}\approx 0.7 \q^{0.2}\mtwo$ \citep{Merritt:06}.

Mass deficits in these models are not related in a simple
way to  the mass in stars ``ejected'' by the binary.
The flux of stars into the binary constitutes a loss term,
$-{\cal F}(E,t)$, on the right hand side of equation~(\ref{eq:dNdt2}),
and in the absence of any other influences, the density of stars
near the center of the galaxy would drop in response to this
term.
Removal of stars also reduces the gravitational force near
the center, contributing to the expansion.
However the second term on the right hand side of equation~(\ref{eq:dNdt2}),
$-\partial F_E/\partial E$, has the opposite effect.
This term represents the change in $N(E,t)$ due to diffusion of 
stars in energy; as the mass deficit increases, so do the gradients
in $f$, which tend to increase the energy flux
and counteract the drop in density.

In principle, these two terms could balance, at least over
some range in energies, allowing the binary to harden
without generating a mass deficit.
This would require
\beq
F_E(E) = \int_E^\infty {\cal F}(E) dE,
\eeq
i.e. the inward flux of stars due to energy diffusion at energy
$E$ must equal the {\it integrated} loss to the binary at
all energies greater than $E$.
However, at sufficiently great distances from the binary,
the relaxation time is so long that the local $F_E(E)$
must drop below the integrated loss term, implying
that the density within this radius will drop.
Growth of a mass deficit reflects the imbalance
between these two terms.

\begin{figure}
\centering
\includegraphics[scale=0.4,angle=-90.]{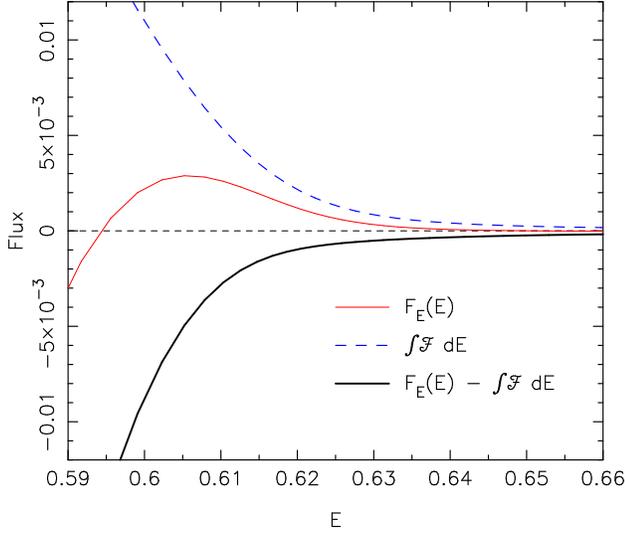}
\caption{Fluxes in the Fokker-Planck integration with
$N=10^{7}$ and $\q=1$, at a time $\sim T_r(r_h)$.
}
\label{fig:fpflux}
\end{figure}

We illustrate this imbalance in Figure~\ref{fig:fpflux}
which shows $F_E(E)$ and $\int{\cal F}(E) dE$
in the Fokker-Planck integration with $\q=1$
at a time $\sim \tr(r_h)$.
The lowest energy in the figure corresponds
roughly to the outer edge of the binary-generated
core.

Yet another mechanism contributes to the growth of mass deficits
in  the Fokker-Planck models.
Even in the absence of the loss term associated with the binary,
the nuclear density profile adopted here for the initial models,
$\rho\sim r^{-0.5}$, implies a ``temperature inversion,''
i.e. a velocity dispersion that increases with radius.
Relaxation drives such a nucleus toward a locally
``isothermal'' form before the onset of core collapse,
causing the central density to drop \citep{Quinlan:96b}.
The binary contributes to this process by maintaining a 
flat density profile near the center, forcing the
temperature inversion to persist.

\section{Implications for Binary Evolution in Galaxies}

Equation~(\ref{eq:tdepend}), based on the Fokker-Planck integrations,
 accurately describes the evolution
of a hard binary in the empty loss cone regime (i.e. in galaxies with
$\mtwo\gap 10^{5.5}\msun$)
given the relaxation time at $r_h$,
while equation~(\ref{eq:trsigma}), based on observed properties of
galactic nuclei (Fig.~\ref{fig:trrh}), provides the mean value of $T_r(r_h)$
for galaxies with black hole mass $\mh=\mtwo$.
Together with equation~(\ref{eq:peters}) for the gravitational
radiation time scale, these relations can be used to
predict mean evolution rates and binary separations in real galaxies
given $(\mh,\q)\equiv (M_2/M_1)$.

\begin{figure}
\centering
\includegraphics[scale=0.45,angle=0.]{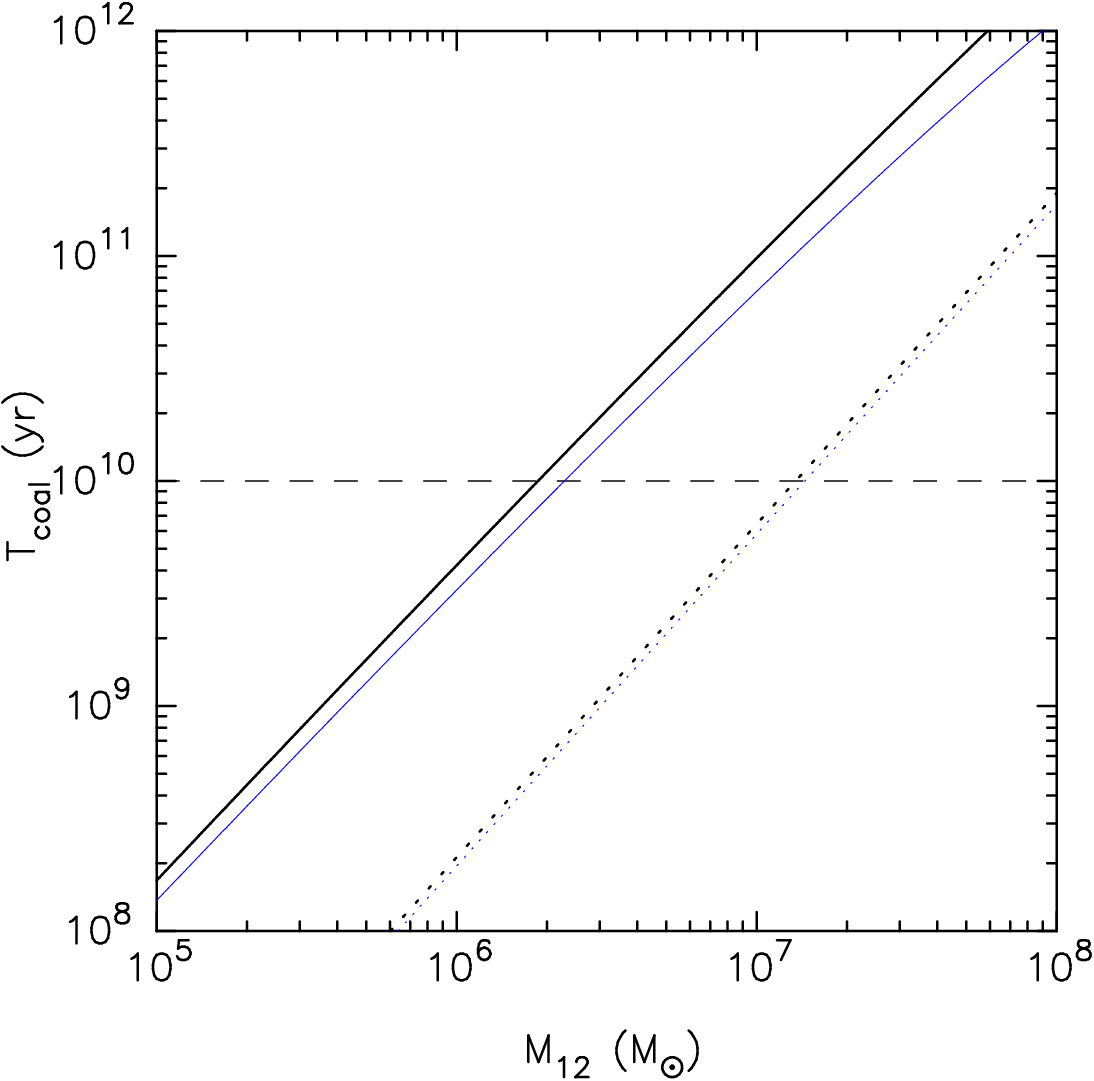}
\caption{Time to coalescence starting from $a=a_h$ as
a function of binary mass.
Solid curves are derived from equation~(\ref{eq:timevsa})
with $y_{max}=\infty$; black/thick: $\q=1$;
blue/thin: $\q=0.1$.
Dotted curves show the evolution time from $a=\aeq$
to $a=0$, i.e. the time spent in the gravitational
radiation regime only.
Equation~\ref{eq:YvsX} gives accurate analytic approximations
to $\tcoal(\mtwo;\q)$.}
\label{fig:tmerge}
\end{figure}

Including the effect of energy lost to gravitational radiation, 
the binary's semi-major axis evolves as
\beq
{d\over dt}\left({1\over a}\right) = 
{d\over dt}\left({1\over a}\right)_{\rm hard} +  
{d\over dt}\left({1\over a}\right)_{\rm gr} 
\eeq
i.e.
\beq
T(a)^{-1} \equiv a{d\over dt}\left({1\over a}\right) = 
\thard^{-1}(a) + \tgr^{-1}(a).
\label{eq:ta}
\eeq
The time for the separation to drop from $a_h$ to $a$ is
\beq
10^{10}{\rm yr}\times \int_0^{y_{max}} {Ay+B\over C+D\left(Ay+B\right)e^{4y}}
\label{eq:timevsa}
\eeq
where
\begin{subequations}
\begin{eqnarray}
C&=&1.25 M_6^{-1.54}, \\
D&=& 1.75\times 10^{-9}\q^{-3}\left(1+\q\right)^6 M_6^{0.65}
\end{eqnarray}
\end{subequations}
and $y_{max}=\ln(a_h/a)$.
The full time to coalescence, $\tcoal$, starting from $a_h$
is given by setting $y_{max}=\infty$
in this expression.
Figure~\ref{fig:tmerge} shows $\tcoal$ as a function of $\mtwo$
for $\q=(1,0.1)$.
Shown separately on this figure is the time to evolve from
$a=\aeq$ to $a=0$, i.e. the time spent in the gravitational radiation
regime alone.
The latter time is a factor $\sim 10$ shorter than the total
evolution time $\tcoal$, which motivates fitting the following 
functional form to $\tcoal(\mtwo;\q)$:
\begin{subequations}
\begin{eqnarray}
Y &=& C_1 + C_2X+C_3X^2, \\
Y&\equiv&\log_{10}\left(\tcoal\over 10^{10}{\rm yr}\right), \\
X&\equiv&\log_{10}\left({\mtwo\over 10^6\msun}\right).
\end{eqnarray}
\label{eq:YvsX}
\end{subequations}
(This functional form  is the integral of equation~\ref{eq:tdepend}.)
A least-squares fit to the curves in Figure~\ref{fig:tmerge} gives
\begin{subequations}
\begin{eqnarray}
\q=1:&&   C_1=-0.372,\ C_2=1.384,\ C_3=-0.025 \\
\q=0.1:&& C_1=-0.478,\ C_2=1.357,\ C_3=-0.041.
\end{eqnarray}
\end{subequations}
The fit of the analytic expressions is better than
$2\%$ ($\q=1$) and $5\%$ ($\q=0.1$); most of the
deviations occur at the high-$\mtwo$ end where coalescence
times are much longer than a Hubble time.

Based on Figure~\ref{fig:tmerge}, binary black holes
would be expected to reach gravitational wave coalescence 
in $10$ Gyr in galaxies with $\mtwo\lap 2\times 10^6\msun$.

\begin{figure}
\centering
\includegraphics[scale=0.45,angle=0.]{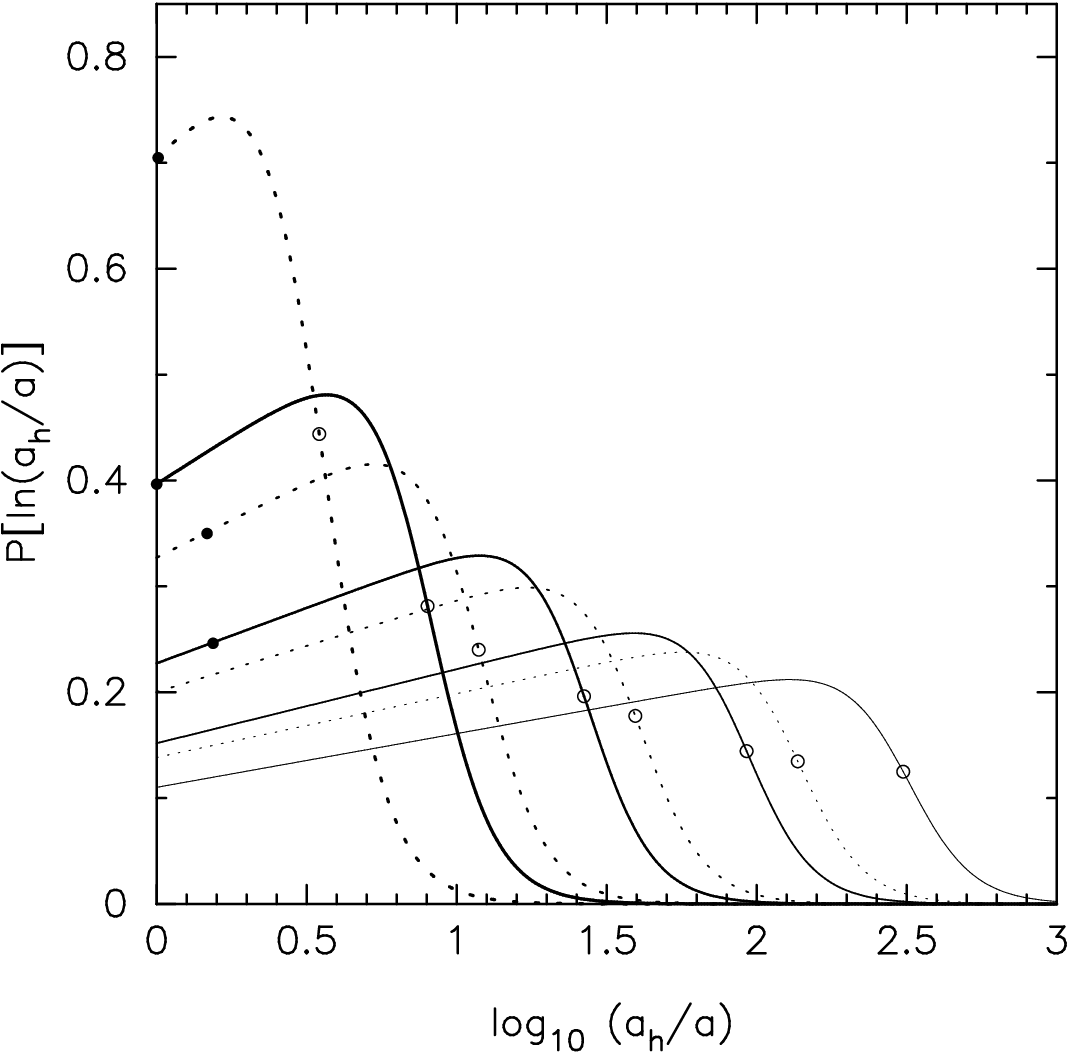}
\caption{Probability of finding a binary black hole in a unit
interval of $\ln a$.
From left to right, curves are for 
$\mtwo=(0.1,1,10,100)\times 10^6\msun$.
Solid(dashed) curves are for $M_2/M_1\equiv\q=1(0.1)$.
Open circles indicate $a=\aeq$; filled circles
correspond to an elapsed time since $a=a_h$ of $10^{10}$ yr.
For the two smallest values of $\mh$, the latter time
occurs off the graph to the right.}
\label{fig:da}
\end{figure}

Figure~\ref{fig:da} shows the probability predicted
by equation~(\ref{eq:ta}) of finding
the binary in a unit interval of $\ln a$,
\beq
P(\ln a)\propto a\left|{da\over dt}\right|^{-1} \propto T(a),
\eeq
for four values of $\mtwo$ and for $\q=(1,0.1)$.
Viewed at a random time before coalescence, a hard
binary is most likely to be seen at $a\approx 2\aeq$,
although the distributions are nearly flat for
$1\le a_h/a \lap 2 a_h/\aeq$.
For $\mtwo\gap 10^7\msun$, evolution for $10$ Gyr
would only bring the binary separation slightly
below $a_h$; in these galaxies the most
likely separation to find a binary would be the
stalling radius \citep{Merritt:06}.

\section{Implications for the Structure of Galaxy Cores}

In the most luminous spheroids, mass deficits generated by 
a binary black hole are likely to persist for the
lifetime of the galaxy, since relaxation times are
much too long for star-star scattering to alter
the phase-space density (cf. Fig.~\ref{fig:trrh}).
In collisional nuclei on the other hand, relaxation times are
short enough that the stellar distribution can be
substantially affected by gravitational encounters
{\it after} the binary black hole has coalesced into 
a single black hole.
A Bahcall-Wolf (1976) cusp will form in a time $\sim T_r(r_h)$
after the binary black hole coalesces into a single hole, 
inside a radius $\sim 0.2 r_h$ \citep{MS:06}.
In addition, the structure of the nucleus beyond
the cusp will continue to evolve, as two-body encounters
drive the stellar ``temperature'' profile
toward isothermality prior to the onset of core collapse
\citep{Quinlan:96b}.
The nuclear density profile at some time after coalescence
will depend on how far along the evolutionary tracks of
Figs.~\ref{fig:0.5_1.0} and~\ref{fig:0.5_0.1} the binary evolved
before coalescing,
as well as on the elapsed time since coalescence.

\begin{figure}
\centering
\includegraphics[scale=0.4,angle=-90.]{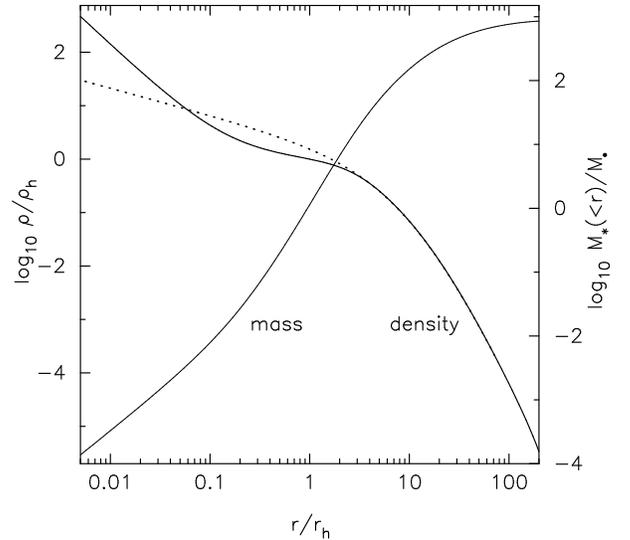}
\caption{Stellar density and mass profiles in a Fokker-Planck
integration with $N=10^9$.
The binary black hole was assumed to coalesce at $t=t_{eq}$
(based on an assumed binary mass of $10^6\msun$)
and the integration was then continued, without the
binary sink term, for one relaxation time at $r_h$.
A Bahcall-Wolf cusp is generated at $r\lap 0.2 r_h$;
the stellar mass within the cusp is $\sim 0.1M_\bullet$.
Dashed line is the initial galaxy model.
}
\label{fig:regen}
\end{figure}

Figure~\ref{fig:regen} illustrates these competing effects
with a concrete example.
A Fokker-Planck integration with $N=10^9$ and $\q=1$
was carried out until a time $t=t_{eq}$; 
$t_{eq}$ was computed as above assuming a binary mass of $10^6\msun$.
The binary was assumed to become a single black hole at
this time; the integration was then continued for a time 
$\tr(r_h)$,
but with the binary loss term $\cal{F}(E)$ set to zero.
As the figure shows, a $\rho\sim r^{-7/4}$ cusp is generated at 
$r\lap 0.2 r_h$.
The net result is a flat core containing at its center a compact
star cluster around the black hole.
The stellar mass within the cusp is $\sim 0.1M_\bullet$.

If binary coalescence were assumed to take place sooner
than $\sim t_{eq}$ (due e.g. to gas-dynamical torques), 
the mass deficit would be
smaller than the value $\sim3.5\mtwo$ generated in this
integration, resulting in a nuclear density profile
more like those of \cite{MS:06}.
As shown in that paper, a regenerated cusp can closely
approximate the (coreless) density profile at the center
of the Milky Way if the elapsed time since binary coalescence
is $\gap 8$ Gyr.
In the integration of Figure~\ref{fig:regen}, on the other hand,
the larger mass deficit is not completely ``erased''
by formation of the cusp.

A nuclear cusp like that in Figure~\ref{fig:regen} would
be unresolved in all but the nearest galaxies.
In fact, recent observations suggest the presence of
compact stellar nuclei (``nuclear star clusters'') at the centers of
most spheroids fainter than $\sim 10^9L_\odot$
\citep{Rossa:06,WH:06,Ferrarese:06}.
The mean mass associated with the nuclei
is a fraction $\sim 0.2\%$ that of the host
galaxy with a $\pm 1\sigma$ range of $0.06\% - 0.52\%$ \citep{Ferrarese:06}.
If we assume that low-luminosity spheroids
contain massive black holes and that the ratio of
black hole mass to spheroid mass is similar to the mean value
$\sim 0.12\%$ characteristic of more luminous galaxies \citep{MF:01},
the  observed nuclei  would have masses that are fractions
$0.5-4$ that of the black holes.
This is somewhat larger than the value $M_{\rm cusp}/\mh\approx 0.1$ 
in the example of Figure~\ref{fig:regen}; on the other hand it is possible
that black holes in faint spheroids carry a larger fraction
of the spheroid mass.
The compact nuclei might also form in very different ways,
e.g. from gas that accumulates at the center.

\section{Alternate Models for Binary Evolution}

\cite{Yu:02} computed evolutionary tracks for binary
black holes at the centers of a sample of early-type
galaxies for which detailed luminosity profiles
were available.
Evolution beyond $a\approx a_h$ was modelled using the
second term on the right hand side of equation~(\ref{eq:dNdt2}),
i.e. the $J$-directed flux of stars into the binary.
The stellar distribution function was assumed fixed in time;
binary-induced changes in the structure of the nucleus 
were ignored, as were changes in stellar energy,
although the time scales associated with both sorts of change
are comparable to the time scale for loss cone repopulation.
As shown here (Figure~\ref{fig:fp}),
allowing for changes in the structure of the nucleus in a Fokker-Planck
model reduces the binary hardening rate by a factor $\sim 2$.
\cite{Yu:02} concluded that binary black holes in spherical
galaxies with $\sigma\lap 90$ km s$^{-1}$ could coalesce in
a Hubble time.
This velocity dispersion corresponds to a binary mass
of $\sim 3.5\times 10^6\msun$ (eq.~\ref{eq:msigma}).
Yu's conclusion is consistent with, but slightly more optimistic
than, the one reached in the current study
(see Fig.~\ref{fig:tmerge}); the differences are probably due 
to Yu's neglect of the back-reaction of the binary on the nucleus.
As in the current study, Yu found a weak dependence
of coalescence time on binary mass ratio.

Some recent studies have inferred rapid 
evolution of supermassive binary black holes 
at the centers of spherical galaxies, 
even in the absence of collisional loss-cone repopulation.
\cite{SHM:07} used detailed three-body scattering experiments to evaluate
the effectiveness of the ``secondary slingshot'' \citep{MM:03}
at extracting energy from massive binaries after they had reached
the stalling radius $a\approx a_h$ in spherical galaxies.
They found that binaries could shrink beyond $a_h$
by factors of $\sim 4 (2)$ for mass ratios of $1 (0.1)$; 
for mass ratios below $\sim 0.01$ 
the secondary slingshot was found to be ineffective.
Almost all of this evolution took place within 
a few galaxy crossing times after the hard binary
had formed; after this time, all of the stars that were
originally within the binary's loss cone had been completely
ejected from the galaxy.

In spite of this very modest evolution,
\cite{SHM:07} concluded that
``even in the absence of other mechanisms driving orbital
decay, pairs involving genuinely supermassive holes 
[i.e. with combined mass $\gap 10^5\msun$]
should not stall''.
This optimistic conclusion appears to have been based on an evaluation of the
mass ``ejected'' by the binary (their Fig. 5), rather than on the more
fundamental criterion of binary separation.
The time to coalescence once a binary reaches the gravitational 
radiation regime is $1/4$ of the time $\tgr$ defined in 
equation~(\ref{eq:tgr});
coalescence occurs in a time of $t_9 $ Gyr if
\beq
{a\over a_h} \approx (0.015,0.034)\times M_{\bullet,6}^{0.16}t_9^{0.25} 
\label{eq:SHM1}
\eeq
where the numbers in parentheses correspond to $\q=(1,0.1)$
respectively. 
The $a/a_h$ values in equation~(\ref{eq:SHM1})
are $\sim 15$ times smaller than those
 found by \cite{SHM:07} after the secondary slingshot had
run its course, implying that the binaries in their model
galaxies would stall at separations far outside the gravitational
radiation regime unless extremely eccentric.

\cite{SHM:07}'s results might still be taken to imply that
massive binaries commence their long-term, relaxation-driven
evolution starting from separations somewhat smaller than
$\sim a_h$, as assumed here.
However such an effect was not apparent in the fully self-consistent
$N$-body simulations of \cite{Merritt:06}.
This is probably due to the neglect by Sesana et al.
of the changes in nuclear structure that accompany 
binary formation.
Sesana et al. computed the initial population of stars 
available to undergo reejections
by assuming a singular isothermal sphere density profile, 
$\rho\propto r^{-2}$, and counting the number of stars on orbits
that intersected the binary.
Even if such a steep density profile were present intially,
it would be converted into a core of much lower density
by the time $a\approx a_h$, and the number of stars available
for the secondary slingshot would be much less than
Sesana et al. estimated.

In the Fokker-Planck integrations presented here (\S5), 
the inclusion of the secondary slingshot had almost no effect on
the long-term behavior of $a(t)$.

\cite{Zier:06a,Zier:06b} also argued that stars near
a binary black hole at the time of its formation
could drive the binary to the gravitational radiation regime
in a very short time.
Zier ignored the secondary slingshot, but assumed that 
a dense cluster of stars would be bound to the binary
at the time that its separation first reached $\sim a_h$.
He found that a cluster having total mass $\sim \mtwo$,
distributed as a steep power-law around the binary,
$\rho\sim r^{-\gamma}$, $\gamma \gap  2.5$, 
could extract enough energy from it via the gravitational
slingshot that $\tgr$ would fall below $10^{10}$ yr.
While no detailed justification for such dense massive
clusters was presented, Zier argued that 
``Low angular momentum matter accumulates in the center''
of merging galaxies, and that 
``Each of the BHs will carry a stellar cusp with a mass of about its
own'' after the merger.
As noted above, recent observations do suggest the presence
of compact nuclei at the centers of low-luminosity spheroids.

$N$-body simulations of Zier's model have yet 
to be carried out, although \cite{MM:01} did follow the evolution
of merging galaxies with initial, $\rho\sim r^{-2}$ cusps
around each of the black holes, close to the value
$\gamma=2.5$ above which Zier infers rapid coalescence.
\cite{MM:01} observed a rapid phase of evolution of
the binary, during which the density cusps were destroyed and
$a$ dropped by a factor of $\sim$ a few below $a_h$.
However this was still far above the separation at which gravitational
radiation would be efficient.

An early, heuristic model for binary evolution was
presented by \cite{Merritt:00} based on the results of the
$N$-body experiments that had been carried out up to that date.
The model assumed that the rate of supply of stars to
the binary was determined by local parameters
(density, velocity dispersion) and was independent
of the nuclear relaxation time.
The model was able to mimic the binary hardening rates
seen in some $N$-body experiments 
(Quinlan \& Hernquist 1997; Chatterjee, Hernquist \& Loeb 2003);
when scaled to real galaxies, it predicted binary coalescence times
that were nearly independent of galaxy mass.
However, the $N$-body results on which the model was based
were subsequently called into question when
they could not be reproduced using more accurate
integrators (Makino \& Funato 2004; Berczik, Merritt \& Spurzem 2005).
M. Volonteri and co-authors adopted the \cite{Merritt:00} 
prescription for binary evolution as a component of their semi-analytic models
of black hole growth \citep{Volonteri:03a,Volonteri:03b,Volonteri:05},
and their models can therefore be expected to substantially over-estimate
the rate of binary evolution in galaxies with $\mh\gap 10^7\msun$.

\section{Summary}

1. Accurate, long-term $N$-body integrations of binary supermassive
black holes at the centers of realistically dense galaxy models
were carried out using particle numbers up to $0.26\times 10^6$.
A new implementation of the Mikkola-Aarseth chain regularization algorithm
was used to treat close interactions involving the black hole particles.
The dependence of the binary's hardening rate on particle number
was quantified by averaging the results of independent integrations.

2. A Fokker-Planck model was developed that includes, for
the first time, changes in the stellar density and potential due to
star-binary interactions.
The Fokker-Planck model was verified by comparison with 
the averaged $N$-body integrations.

3. Based on the Fokker-Planck integrations and on empirical scaling
relations, binary evolution in real galaxies was shown to take place in the 
``empty loss cone'' (diffusive) regime for binaries with total mass above 
about $10^{5.5}\msun$.
This regime is out of range of particle numbers currently
feasible via direct $N$-body simulation but can
be efficiently treated via the Fokker-Planck approximation.

4. Accurate analytical expressions were derived that 
reproduce the predictions of the Fokker-Planck model for
the time-dependence of binary semi-major axis (equation~\ref{eq:tdepend})
and the time to coalescence (equation~\ref{eq:YvsX})
in the diffusive regime.

5. Based on the Fokker-Planck integrations and on empirical scaling relations, 
gravitational-radiation coalescence will occur in
$10$ Gyr or less for galaxies with binary masses
$\lap 2\times 10^6\msun$ or central 
velocity dispersions $\lap 80$ km s$^{-1}$; the coalescence time
depends only weakly on binary mass ratio
(Fig.~\ref{fig:tmerge}).
Binaries with masses $\gap 10^7\msun$ will remain stalled
for a Hubble time.

6. A core, or ``mass deficit,'' is created 
as a result of competition between ejection of stars by the binary 
and re-supply of depleted orbits via gravitational (star-star)
encounters.
Mass deficits as large as $\sim 4(M_1+M_2)$ were found to be
generated before 
coalescence (Fig.~\ref{fig:0.5_1.0},\ref{fig:0.5_0.1}).

7. After the black holes coalesce, 
a Bahcall-Wolf cusp forms around the
single hole in approximately one relaxation time, resulting in
a nuclear density profile with a flat core and an inner, compact 
cluster (Fig.~\ref{fig:regen}), similar to what is
observed at the centers of low-luminosity spheroids.

\bigskip\bigskip

This work was supported by grants 
AST-0206031, AST-0420920 and AST-0437519 from the 
NSF, grant 
NNG04GJ48G from NASA,
and grant HST-AR-09519.01-A from
STScI.

\end{document}